\documentclass[aps,pre,preprint]{revtex4-2}
\usepackage{amsmath}
\usepackage{graphicx}
\usepackage{bbm}
\usepackage{subfigure}
\usepackage[caption=false]{subfig}
\begin{document}
\title{Collective Quantum Approach to Surface Plasmon Resonance Effect}
\author{M. Akbari-Moghanjoughi}
\affiliation{Faculty of Sciences, Department of Physics, Azarbaijan Shahid Madani University, 51745-406 Tabriz, Iran\\ massoud2002@yahoo.com}

\begin{abstract}
In this research we present a theory of the surface plasmon resonance (SPR) effect based on the dual length-scale driven damped collective quantum oscillations of the spill-out electrons in plasmonic material surface. The metallic electron excitations are modeled using the Hermitian effective Schr\"{o}dinger-Poisson system, whereas, the spill-out electron excitations are modeled via the damped non-Hermitian effective Schr\"{o}dinger-Poisson system adapted appropriately at the metal-vacuum interface. It is shows that, when driven by external field, the system behaves like the driven damped oscillator in wavenumber domain, quite analogous to the driven damped mechanical oscillation in frequency domain, leading to the collective surface spill-out electron excitation resonance. In this model the resonance occurs when the wavenumber of the driving pseudoforce matches that of the surface plasmon excitations which can be either due to single-electrons or collective effects. Current theory of SPR is based on longitudinal electrostatic excitations of the surface electrons, instead of the polariton excitation parallel to the metal-dielectric or metal-vacuum surface. Current theory may also be extended to use for the localized surface plasmon resonance (LSPR) in nanometer sized metallic surfaces in non-planar geometry. A new equation of state (EoS) for the plasmon electron number density in quantum plasmas is obtained which limits the plasmonic effects in high-density low-temperature electron gas regime, due to small transition probability of electrons to the plasmon energy band.
\end{abstract}

\date{\today}

\maketitle
\newpage

\section{Introduction}

Plasmonics is a rapidly developing fundamental subject of research which combines the theory and applications spanning from physics, chemistry, engineering, biology, medicine and many other areas of active interdisciplinary fields \cite{atw,man1,maier}. Plasmonics had a long tradition since the discovery of surface plasmon polariton in 1950's. It has attracted significant attention after the discovery of the surface-enhanced Raman scattering (SERS) effect in 1970's \cite{xu} and continued to grow exponentially during the last decade, due to explosive technological developments in nanofabrication and low dimensional miniaturized semiconductor industry \cite{haug,mark}. Many new fields of research, such as nanoplasmonics, nanophotonics \cite{seeger,hu}, nano-optics \cite{yofee} are emerging around the subject because of evergrowing applications of the plasmonics in futuristic nanotechnology, ultrahigh sensitivity and terahertz-range dielectric response of electrons to environmental changes. The surface plasmon resonance (SPR) is versatile technology used in determining particular substances called analytes. SPR biosensors are also used in label-free detection of various important biomarkers \cite{zhu}. Despite a huge experimental advancement \cite{jian,fey1,hugen,sun1,sun2,sun3} and success in the field, the theoretical developments remain classical. The best investigation tools for consistent results in plasmonic phenomena are numerical simulation and computational algorithms, which are time consuming due to many body complexity from one hand and prone to error, on the other. Furthermore, the classical Drude model and numerical simulations do not provide fundamental insight into the underlying atomic-scale physical phenomena at the quantum level.

Plasmonic systems, on the other hand, provide variety of interesting optical properties depending on many parameters, such as, the shape, size, periodicity and the material combination. The extreme sensitivity of electron oscillations on material configurations provides a great chance for industry technicians to invent vast variety of electronic devices by manipulation of involved parameters. Plasmonic devices are also main components of biosensors, optical modulators, photovoltaic technology etc. There are, however, technological obstacle and insufficient theoretical developments which limit the harvesting capacity of the plasmonic effects.  By overcoming these limitations, plasmonic devices can find their full potential in optical emitters \cite{and}, plasmon focusing \cite{stock}, nanoscale waveguiding \cite{qui}, optical antennas \cite{muh}, communication device \cite{umm}, solar cells \cite{zhuo}, plasmonic sensors \cite{goy}, modulator \cite{haf} nanoscale swiches \cite{kar} and spasers \cite{oul}. Moreover, collective electron excitations may become the best candidates for developing effective hot electron generation \cite{cesar} for solar energy conversion towards improved photovoltaic and catalytic designs via the collective energy transport phenomena. The new plasmonic energy conversion may require several important considerations regarding the plasmonic materials, integrated architecture and device fabrication methods \cite{jac}. Recent investigations reveal that guiding and intensification of light by employing the right plasmonic geometries leads to more effective solar-cell designs in which light is more efficiently absorbed in a single quantum well, or in a single layer of quantum dots or molecular chromophores \cite{at2}.

Plasmonics is also an appealing theoretical field of research, since, fundamental electronic excitations control almost every aspect of a solid, such as conductivity, heat and electrical transport \cite{kit}. The electrons uniquely interact with the lattice periodicity ans electromagnetic field to produce complex energy band structure which determines the thermodynamic behavior of a solid \cite{ash}. However, understanding the fundamental plasmonic behavior of interacting electron gas requires the improvement of advanced quantum many body \cite{fetter,mahan} and quantum electrodynamic \cite{fey} theories. Although, the foundation of the theory has been laid by previous pioneering works \cite{fermi,bohm,bohm1,bohm2,pines,levine,klim,5,6,7,8,9,50,51,52,53,lind}, the quantum mechanical treatment of electron system are usually based on various limiting approximations, such as, the single-electron energy dispersion and random-phase approximation. Due to complex nature of plasmonic phenomenon and the delicate interplay between the individual electrons and their collective excitation, some important aspects of the phenomenon are underestimated when the single-electron and collective behavior of the electron system are not treated in equal footing. For instance, there has been an intense recent debate \cite{bonitz1,sea1,bonitz2,sea2,bonitz3,akbarihd,sm,michta,moldabekov} on the inconsistent prediction of attractive force between static screened ions in quantum plasmas between the linearized quantum hydrodynamic \cite{gard,ichimaru1,ichimaru2,ichimaru3,man0,haas1,moldbon} and density functional theory (DFT) \cite{kohn} treatments . While these theories can be reconciled when corrections are taken into account \cite{elak}, the fundamental theory of plasmonics must go well beyond these correction and approximations. In an unpublished resent research the Lindhard theory is used with a more general dual length-scale energy dispersion which includes both single-electron as well as collective behavior \cite{akblind}. It is seen that, the obtained results of such adjustment deviates substantially from prediction of both conventional quantum hydrodynamic and the Lindhard theory. The quantum hydrodynamic theory which is based on the Wigner-Poisson system has shown many success in predicting collective phenomena in dense electron environments \cite{markl,shuk,brd,se,sten,ses,brod1,mark1,man3,kim}. However, uses the independent single-electron Hamiltonian to generalize to many body effects and therefore ignores the energy band structure effect due to the electrostatic interactions. A more elaborative treatment in a recent research the Schr\"{o}dinger-Poisson model \cite{fhaas} has been used to obtain the generalized energy dispersion relation from which collective phenomena is studied in both single-electron and collective scales at the same time. Some interesting new collective quantum effects have been predicted within the Schr\"{o}dinger-Poisson model \cite{akbquant,akbheat,akbdual,akbedge,akbint,akbnat1,akbnat2}. In this research we develop a new theory of SPR based on the Schr\"{o}dinger-Poisson model based on the driven excitations in half-space collective quantum excitations. The paper is organized in the following manner. The Linear quantum excitations are modeled in Sec. II. The model for spill-out electron excitations is provided in Sec. III. Driven collective excitations at the surface is considered in Sec. IV and conclusions are drown in Sec V.

\section{Linear Collective Quantum Excitations}

The elementary collective quantum excitations of the electron gas can be described using various modeling such as kinetic, density functional and hydrodynamic theories. Here we use the effective Schr\"{o}dinger-Poisson system, within the framework of quantum hydrodynamic model \cite{haasbook}, in order to study plasmonic excitations in vacuum-metal-vacuum configuration. The one-dimensional collective dynamic state of the electrostatically interacting metallic electrons follow the Hermitian coupled pseudoforce system \cite{akbnat2}
\begin{subequations}\label{sp}
\begin{align}
&i\hbar \frac{{\partial N({\bf{r}},t)}}{{\partial t}} =  - \frac{{{\hbar ^2}}}{{2m}}\Delta N({\bf{r}},t) - e\phi ({\bf{r}})N({\bf{r}},t) + \mu N({\bf{r}},t),\\
&\Delta \phi ({\bf{r}}) = 4\pi e\left[ {|N({\bf{r}}){|^2} - {n_0}} \right],
\end{align}
\end{subequations}
in which ${\cal{N}}({\bf{r}},t)=\psi({\bf{r}},t)\exp[iS({\bf{r}},t)/\hbar]$ denotes the collective statefunction, in combination with the electrostatic potential $\phi(x)$, characterizing the probability density of the electron system which is related to the local electron number density via the definition, $n({\bf{r}})=\psi({\bf{r}})\psi^*({\bf{r}})$ in which $n_0$ denotes the neutralizing positive background charge density in the metallic slab region of length $L$ and the electron fluid momentum is given by, ${\rm{p(r,t) = }}\nabla {\rm{S(r,t)}}$. The electrostatic potential $\phi({\bf{r}})$, plays the essential role of coupling the Schr\"{o}dinger equation to the Poisson's relation.

In the slow electron gas compression limit, the chemical potential $\mu$ is related to the electron number-density via the isothermal equation of state (EoS) \cite{elak}
\begin{subequations}\label{np}
\begin{align}
&{n_e(\mu,T)} = \frac{{{2^{1/2}}m{^{3/2}}}}{{{\pi ^2}{\hbar ^3}}}  \int_{0}^{ + \infty } {\frac{{\sqrt{{\varepsilon}} d{\varepsilon}}}{{{e^{\beta ({\varepsilon}-\mu)}} + 1}}},\\
&{P_e(\mu,T)} = \frac{{{2^{3/2}} m{^{3/2}}}}{{3{\pi ^2}{\hbar ^3}}}\int_0^{ + \infty } {\frac{{{{\varepsilon}^{3/2}} d{\varepsilon}}}{{{e^{\beta ({\varepsilon} - {\mu})}} + 1}}.}
\end{align}
\end{subequations}
where $\beta=1/k_B T$ with $T$ being the equilibrium electron temperature and $P_e$ the quantum statistical electron gas pressure. Note that the simple thermodynamic identity, $n_e\nabla\mu=\nabla P_e(n_e)$, holds.

For the one dimensional model in the stationary limit of equilibrium state, $p=0$, the wavefunction $\psi(x,t)$ may be decomposed into separate variables leading to following linearized system, after appropriate normalization of variables \cite{akbdual}
\begin{subequations}\label{pf}
\begin{align}
&i\hbar \frac{{d\psi_1(t)}}{{dt}} = \varepsilon\psi_1(t),\\
&\frac{{d\Psi_1 (x)}}{{dx}} + \Phi_1 (x) + E\Psi_1 (x) = 0,\\
&\frac{{d\Phi_1 (x)}}{{dx}} - \Psi_1 (x) = 0,
\end{align}
\end{subequations}
where $\mu_0$ is the equilibrium chemical potential and $\varepsilon$ is the collective quantum-state energy eigenvalue. Note that the linear expansion scheme, $\{\psi^0=1,\phi^0=0,\mu^0=\mu_0\}$ is employed to normalize the functionals $\Psi({\bf{r}})=\psi({\bf{r}})/n_0$ and $\Phi({\bf{r}})=e\phi({\bf{r}})/E_p$ and variables $x=x/l_p$ and $t=t\omega_p$ in which $\omega_p=\sqrt{4\pi e^2 n_0/m}$ is the conventional plasmon frequency, $l_p=1/k_p$ is the characteristic plasmon length with $k_p=\sqrt{2 m E_p}/\hbar$ being the coresponding plasmon wavenumber with $E_p=\hbar\omega_p$ being the plasmon energy. The normalized energy $E=\epsilon - \sigma$ with $\epsilon=\varepsilon/E_p$ and the $\sigma=\mu_0/E_p$ denotes the energy scale measured from the Fermi energy level, $E_F=\hbar^2(3\pi n_0)^{2/3}/2m$, of the degenerate electron gas. The Fourier analysis of coupled pseudoforce system (\ref{pf}) admits the generalized matter wave energy dispersion, $E=k^2+1/k^2$, with the first and second terms characterizing the single-electron and collective excitations, respectively. The presence of electrostatic coupling between electrons leads to the dual length-scale plasmonic excitations of the electron gas.

The real-valued solutions to the time-independent system (\ref{pf}) follows
\begin{subequations}\label{solpf}
\begin{align}
&{\Phi _1}(x) = \frac{1}{\alpha }\left[ {\left( {{c_1}k_{21}^2 + {c_3}} \right)\cos ({k_{11}}x) - \left( {{c_1}k_{11}^2 + {c_3}} \right)\cos ({k_{21}}x)}\right]\\
&+ \frac{2}{\alpha }\left[ {\frac{{\left( {{c_2}k_{21}^2 + {c_4}} \right)\sin ({k_{11}}x)}}{{{k_{11}}}} - \frac{{\left( {{c_2}k_{11}^2 + {c_4}} \right)\sin ({k_{21}}x)}}{{{k_{21}}}}} \right].\\
&{\Psi _1}(x) = 1 + \frac{1}{\alpha }\left[ {\left( {{c_1} + {c_3}k_{21}^2} \right)\cos ({k_{21}}x) - \left( {{c_1} + {c_3}k_{11}^2} \right)\cos ({k_{11}}x)} \right]\\
&+ \frac{2}{\alpha }\left[ { {\frac{{\left( {{c_2} + {c_4}k_{21}^2} \right)\sin ({k_{21}}x)}}{{{k_{21}}}} - \frac{{\left( {{c_2} + {c_4}k_{11}^2} \right)\sin ({k_{11}}x)}}{{{k_{11}}}}}} \right],
\end{align}
\end{subequations}
where $c_i$ are the integration constants characterizing the boundary conditions with $\alpha=\sqrt{E^2-4}$, $k_{11}=\sqrt{(E-\alpha)/2}$ and $k_{21}=\sqrt{(E+\alpha)/2}$ being wavenumbers of the collective and single-electron excitations which satisfy the complementarity-like relation, $k_{11}k_{21}=1$.

\begin{figure}[ptb]\label{Figure1}
\includegraphics[scale=0.6]{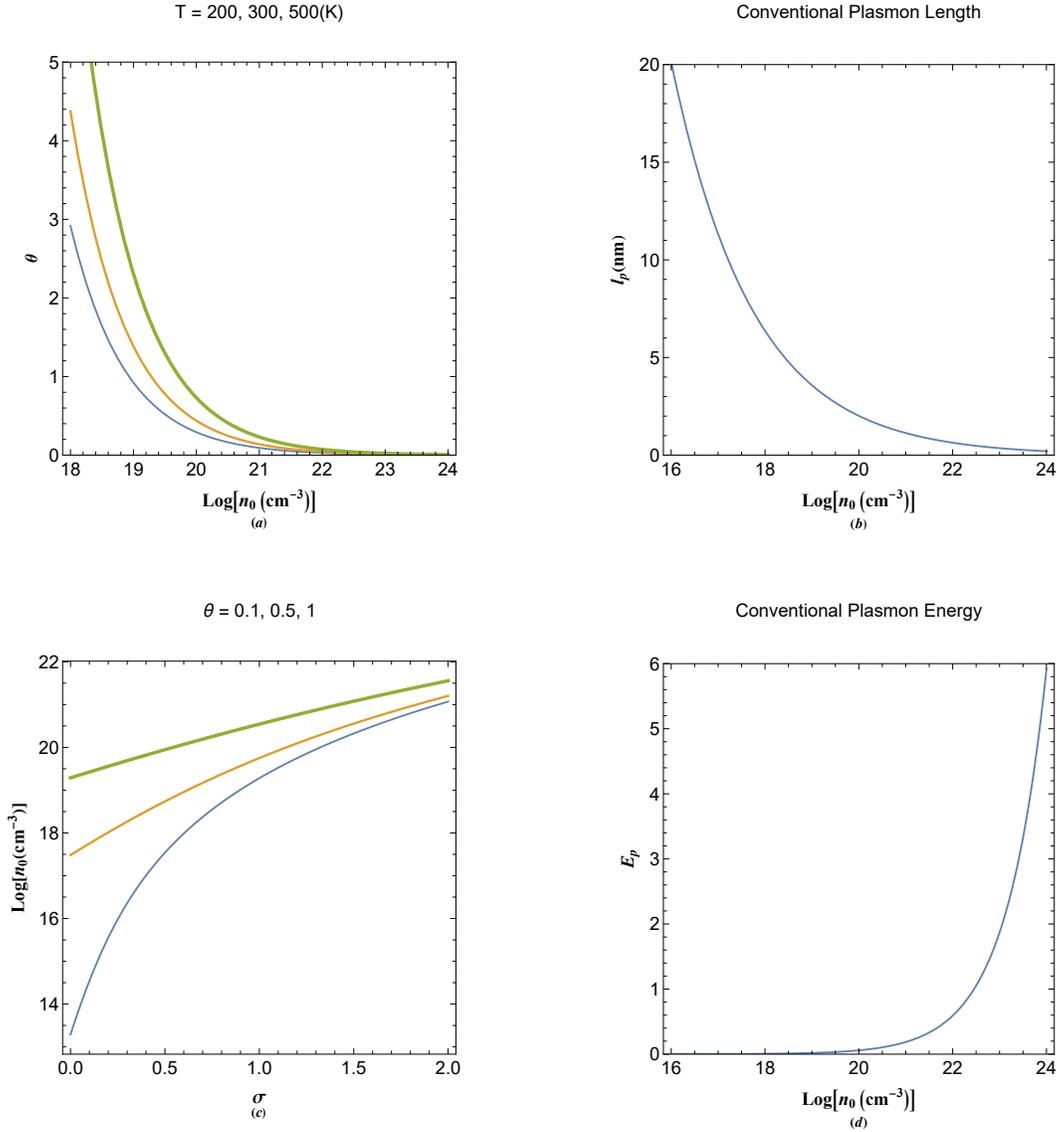}\caption{(a) The variation in normalized plasmon temperature, $\theta$, with electron concentration, for different values of the electron temperature. (b) Variation of the plasmon length with the electron number density. (c) Variation of normalized chemical potential $\sigma=\mu_0/E_P$ with free electron concentration, for different values of the normalized plasmon temperature. (d) Changes in the plasmon energy of the electron gas with electron concentration. The increase in thickness of curves in plots (a) and (c) refer to increases in the varied parameter above given frames.}
\end{figure}

Figure 1 shows variations in various plasmon parameters in terms of the electron number density. Fig 2(a) depicts the variation of normalized electron temperature, $\theta=T/T_p$, where $T_p=E_p/\hbar$ is the characteristic plasmon temperature with electron concentration for different values of the electron temperature, $T$. It is noted that the plasmon temperature, $T_p$, despite its deceptive name has noting to do with real temperature and analogous to the Fermi temperature depends solely on the electron number density of the electron gas. It is remarked that $\theta$ decreases sharply with increase in electron number density being larger for larger temperature values. Variation the plasmon length $l_p$ is shown in Fig. 2(b) in terms of the electron concentration. The plasmon length is seen to also decreases with increase in the electron density with typical values of few tenth of a nanometer in metallic electron density range. Figure 2(c) depicts variation in the normalized chemical potential with electron concentration for different values of $\theta$. The normalized chemical potential, $\sigma$ only varies greatly with electron number density for lower $\theta$ values. Figure 2(d) shows the dependence of plasmon energy on the electron number density. This parameter in the fully degenerate electron gas regime has typical values of few electron Volts and increases sharply as the electron number density is increased.

An important note is to be made here regarding the standard definitions of plasmon parameters in degenerate electron regime. Conventionally, the plasmon frequency, hence, the plasmon energy is defined in terms of the equilibrium number density of free electrons. However, at low temperature or equivalently high electron number density regime, majority of metallic electrons are packed below the Fermi level which prohibits them from collective contributions. It will be shown that, in order to correctly describe the plasmonic phenomena in metallic regime, these parameter definitions must be altered according to the plasmon number density, $n_p$ based on transition probability of electrons to the so-called plasmon energy band. According to the new energy band definition for collective excitations, one is able to compute the spill-out electron density at the surface of metals and nanoparticles, which is considered in the following section.

\section{Spill-out Electron Excitations}

The spill-out electrons in region 2 has been recently modeled through the following non-Hermitian pseudoforce system \cite{akbedge}
\begin{subequations}\label{pfn}
\begin{align}
&\frac{{{d^2}\Psi_2 (x)}}{{d{x^2}}} + 2\kappa \frac{{d\Psi_2 (x)}}{{dx}} + \Phi_2 (x) + E\Psi_2 (x) = 0,\\
&\frac{{{d^2}\Phi_2 (x)}}{{d{x^2}}} + 2\kappa \frac{{d\Phi_2 (x)}}{{dx}} - \Psi_2 (x) = 0,
\end{align}
\end{subequations}
in which $\kappa$ is the normalized damping parameter characterizing the spacial decay of the collective statefunctions in vacuum side of the metallic sandwich. Although, this parameter may have a complex dependence on other parameters, it can be directly measured using a Langmuir probe. This damping parameter plays the analogous role of the tunneling parameter in the single-electron quantum system. The only difference in this case is that a collective tunneling of electrons leading to the electron spill-out effect has also an oscillatory density behavior along the damping character. Different scenarios of the damping are considered in the Ref. \cite{akbdual}. Recently it has been shown that electrostatic interactions among the spill-out electrons in combination of the single-electron excitations leads to a new photo-plasmonic phenomenon and the collective tunneling of the electrons on the irradiated metallic surface though the photo-plasmonic energy level \cite{akbnat1}.

The real-valued solutions to the coupled non-Hermitian system (\ref{pfn}) follows
\begin{subequations}\label{solpfn}
\begin{align}
&{\Phi_2}(x) = \frac{{{{{e}}^{ - \kappa x}}}}{\alpha }\left[ {\left( {{d_1}k_{21}^2 + {d_3}} \right)\cos ({k_{22}}x) - \left( {{d_1}k_{11}^2 + {d_3}} \right)\cos ({k_{12}}x)} \right]\\
& + \frac{{{{{e}}^{ - \kappa x}}}}{{\alpha {k_{12}}}}\left[ {\left( {\alpha  - E} \right)\left( {{d_1}\kappa  + {d_2}} \right) - 2\left( {{d_3}\kappa  + {d_4}} \right)} \right]\sin ({k_{12}}x)\\
& + \frac{{{{{e}}^{ - \kappa x}}}}{{\alpha {k_{22}}}}\left[ {\left( {\alpha  + E} \right)\left( {{d_1}\kappa  + {d_2}} \right) + 2\left( {{d_3}\kappa  + {d_4}} \right)} \right]\sin ({k_{22}}x).\\
&{\Psi_2}(x) = \frac{{{{{e}}^{ - \kappa x}}}}{\alpha }\left[ {\left( {{d_1} + {d_3}k_{21}^2} \right)\cos ({k_{22}}x) - \left( {{d_1} + {d_3}k_{11}^2} \right)\cos ({k_{12}}x)} \right]\\
& + \frac{{2{{{e}}^{ - \kappa x}}}}{{\alpha {k_{22}}}}\left[ {\kappa {d_1} + {d_2} + k_{21}^2\left( {\kappa {d_3} + {d_4}} \right)} \right]\sin ({k_{22}}x)\\
& - \frac{{2{{{e}}^{ - \kappa x}}}}{{\alpha {k_{12}}}}\left[ {\kappa {d_1} + {d_2} + k_{11}^2\left( {\kappa {d_3} + {d_4}} \right)} \right]\sin ({k_{12}}x),
\end{align}
\end{subequations}
where $d_i$ are integration constants set by the boundary conditions, $k_{12}=\sqrt{k_{11}^2-\kappa^2}$ and $k_{22}=\sqrt{k_{21}^2-\kappa^2}$ are the characteristic generalized wavenumbers of the spill-out electron excitations admitting the new energy dispersion relation, $E=(k^2+\kappa^2)+1/(k^2+\kappa^2)$, including the pseudo-damping effect. The complementarity-like relation in this case it, $k_{12}k_{22}=\sqrt{\kappa^4-E\kappa^2+1}$ which reduces to the undamped plasmon energy dispersion in the limit $\kappa=0$.

The plasmonic gas has some distinct thermodynamic properties which can be obtained quite analogous to the free electron gas. The number density of plasmon mode is obtained via, $N(k)=4\pi k^3/3$, resulting in the plasmon density of states (DoS), $g(k)=(dN/dk)/|dE/dk|$. Given that the electron occupation function in plasmon mode is dictates the Pauli exclusion we have, $F(k,\kappa,\theta)=1/[1+\exp(E/\theta)]$ in normalized form. Therefore, the thermodynamic quantities such as the normalized plasmon electron number-density $n_p(\theta,\kappa)$, the internal energy of plasmon electrons, $U_p(\theta,\kappa)$ and heat capacity contribution from plasmonic excitations, $c_p(\theta,\kappa)$, are given as
\begin{subequations}\label{ther}
\begin{align}
&{n_p}(\theta ,\kappa ) = \int\limits_0^\infty  {\frac{{g(k,\kappa ){d_k}E(k,\kappa )dk}}{{1 + \exp [E(k,\kappa )/\theta ]}}},\hspace{3mm}g(k,\kappa ) = \frac{{2\pi k{{\left( {{k^2} + {\kappa ^2}} \right)}^2}}}{{\left| {{{\left( {{k^2} + {\kappa ^2}} \right)}^2} - 1} \right|}},\\
&U_p(\theta,\kappa) = \int\limits_0^\infty  {\frac{{{g(k,\kappa)}{E(k,\kappa)}{d_k}{E(k,\kappa)}dk}}{{1 + \exp [{E(k,\kappa)}/\theta]}}},\hspace{3mm}{E(k,\kappa)} = (k^2+\kappa^2)+1/(k^2+\kappa^2),\\
&c_p(\theta ,\kappa ) = \int\limits_0^\infty  {g(k,\kappa )E(k,\kappa ){d_\theta }F(k,\kappa,\theta){d_k}E(k,\kappa )dk},
\end{align}
\end{subequations}
As previously noted, there is a fundamental difference between plasmon electron number density, $n_p$, and that of the free electrons, $n_0$, appearing in (\ref{ther}) and the Fermi energy, respectively. The former quantity refers to the excited electrons to the plasmon level, whereas, the later is the total electron number density at the equilibrium temperature. In the fully degenerate quantum plasmas only fraction of free electrons below the Fermi energy are excited to the plasmonic level, depending on the plasma density-temperature regime and contribute to collective oscillations.

\begin{figure}[ptb]\label{Figure2}
\includegraphics[scale=0.6]{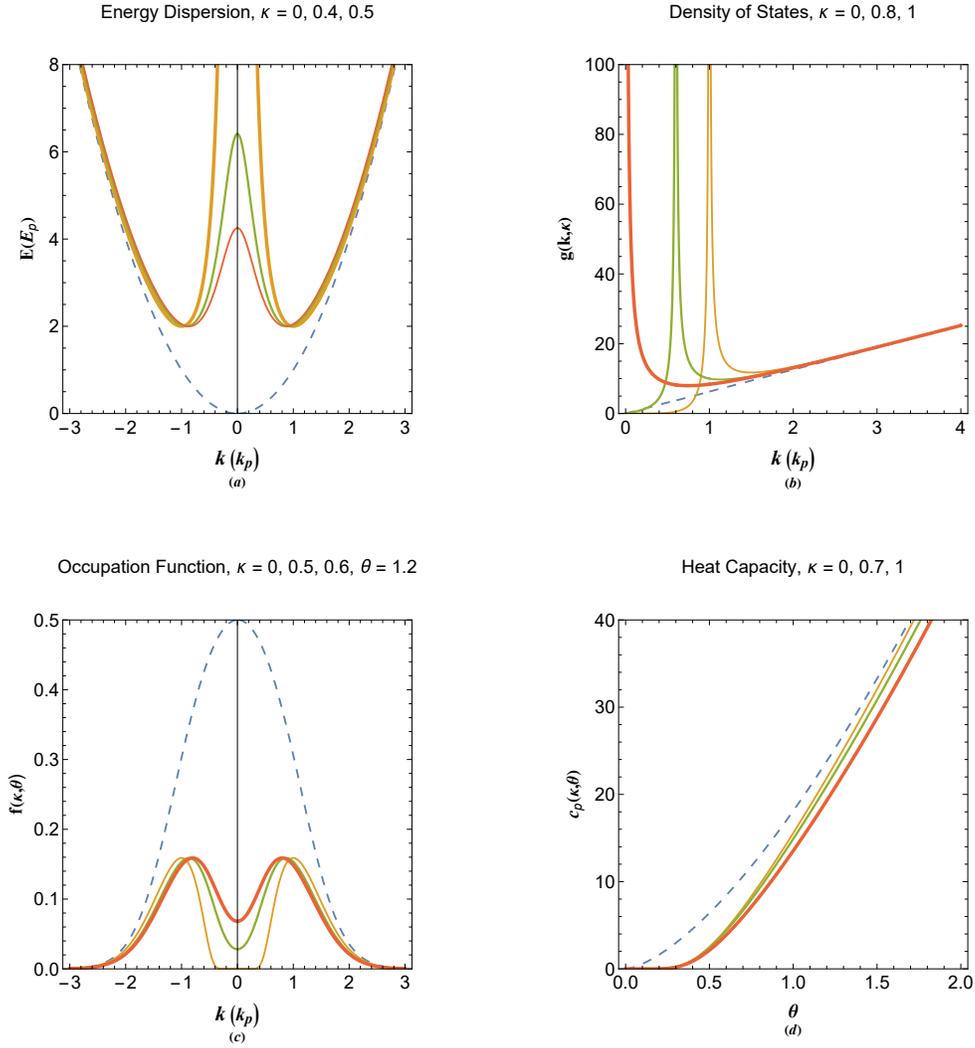}\caption{(a) The plasmon energy dispersion profiles for different damping parameter values. (b) The plasmon density of states (DoS) for different values of the damping parameter, $\kappa$. (c) The plasmon energy band occupation function for different values of the damping parameter. (d) The plasmon heat capacity contribution for different values of $\kappa$. The dashed curves indicate the corresponding curves for the free electron model. The increase in thickness of curves in plots refer to increases in the varied parameter above given frames.}
\end{figure}

In Fig. 2 we have shown the variations of thermodynamic quantities corresponding to electron which are excited to the plasmon energy band. The generalized plasmon energy dispersion for different values of the damping parameter, $\kappa$, is depicted in Fig. 3(a). The dashed curve corresponds to the free electron energy dispersion. For undamped case, $\kappa=0$, the plasmon conduction band takes place at, $k=1$, (in plasmon wavenumber unit), with an energy band gap for $0<E<2$ (in plasmon energy unit) in which the plasmon excitations are unstable. For values of $E(k,0)>2$ the collective excitations are of dual-tone nature, while, for $E\simeq 2$ the quantum beating behavior takes place for which the collective and single-electron excitation scale-lengths match approximately. The dispersion curve consists of two distinct branches joining at $k=1$ with the left/right branch corresponding to the wave-like/particle-like excitations. For the damped plasmons, $\kappa\ne 0$, (as the case for spill-out electrons) the wave-like oscillations become unstable above a critical wavenumber value, $k_c=\kappa^2+1/\kappa^2$ and only single-electron excitations are allowed. However, the damping parameter does not significantly affect the particle-like oscillation branch. The plasmon electron DoS is shown in Fig. 2(b) for different values of the damping parameter. The dashed line corresponds to the free electron model. There is a Van-Hove-like singularity at the critical wavenumber value, $\sqrt{1-\kappa^2}$ which moves to lower wavenumbers with increase of the damping parameter. This feature is quite analogous to the well-known Van-Hove singularity in the DoS at the Fermi level of most solids \cite{kit,ash}. It is clearly remarked that, at the small wavelength limit the plasmon electron DoS approaches the value of the free electron gas. The later is because at such limit the single electron excitations dominate the collective ones. Figure 2(c) depicts the plasmon occupation number showing a peak which coincides with $k=1$ for $\kappa=0$, with the dashed curve corresponding to the free electron occupation function peaking at $k=0$. The plasmon occupation number peak position, however, moves to lower $k$-values, as the damping parameter is increased and for the total damping $\kappa=1$ the occupation number peaks at $k=0$. Note that the long wavelength plasmon modes become more occupied as the damping parameter increases and looks like the single-electron dashed curve as expected due to domination of these excitations. The variation of plasmonic heat capacity is shown in Fig. 2(d) for different values of the damping parameter in terms of the normalized temperature. The corresponding single-electron heat capacity (dashed curve) shows increased contribution at lower temperature because the low temperature limits the electron transitions to plasmon band are greatly diminished. It is also remarked that plasmon damping has significant effect on the plasmon density only at higher electron temperatures.

\section{Surface Plasmon Resonance Effect}

In order to study the surface plasmon resonance effect, we consider a more generalized driven coupled pseudoforce system
\begin{subequations}\label{pfnd}
\begin{align}
&\frac{{{d^2}\Psi_2 (x)}}{{d{x^2}}} + 2\kappa \frac{{d\Psi_2 (x)}}{{dx}} + \Phi_2 (x) + E\Psi_2 (x) = F \cos(q x),\\
&\frac{{{d^2}\Phi_2 (x)}}{{d{x^2}}} + 2\kappa \frac{{d\Phi_2 (x)}}{{dx}} - \Psi_2 (x) = 0,
\end{align}
\end{subequations}
in which $F$ and $q$ are, respectively, the pseudoforce driving normalized amplitude and wavenumber. The real-valued complementary solutions of the system (\ref{pfnd}) reads
\begin{subequations}\label{pfndsol}
\begin{align}
&\Phi (x) = {\Phi _2}(x) - \frac{{F{e^{ - \kappa x}}}}{{\alpha {\eta _1}{\eta _2}}}\left[ {{\eta _1}\left( {{q^2} - k_{21}^2} \right)\cos ({k_{22}}x) - {\eta _2}\left( {{q^2} - k_{11}^2} \right)\cos ({k_{12}}x)} \right]\\
& + \frac{{F\kappa {e^{ - \kappa x}}}}{{\alpha {\eta _1}{\eta _2}}}\left[ {{\eta _1}\left( {{q^2} + k_{21}^2} \right)\frac{{\sin ({k_{22}}x)}}{{{k_{22}}}} - {\eta _2}\left( {{q^2} + k_{11}^2} \right)\frac{{\sin ({k_{12}}x)}}{{{k_{12}}}}} \right]\\
& + \frac{F}{{{\eta _1}{\eta _2}}}\left\{ {\left[ {\left( {{q^2} - k_{11}^2} \right)\left( {{q^2} - k_{21}^2} \right) - 4{q^2}{\kappa ^2}} \right]\cos (qx) + 2q\kappa \left( {k_{11}^2 + k_{21}^2 - 2{q^2}} \right)\sin (qx)} \right\},\\
&\Psi (x) = {\Psi _2}(x) + \frac{{F{{{e}}^{ - \kappa x}}}}{{\alpha {\eta _1}{\eta _2}}}\left[ {{\eta _1}k_{21}^2\left( {{q^2} - k_{21}^2} \right)\cos ({k_{22}}x) - {\eta _2}k_{11}^2\left( {{q^2} - k_{11}^2} \right)\cos ({k_{12}}x)} \right]\\
& - \frac{{F\kappa {e^{ - \kappa x}}}}{{\alpha {\eta _1}{\eta _2}}}\left[ {{\eta _1}k_{21}^2\left( {{q^2} + k_{21}^2} \right)\frac{{\sin ({k_{22}}x)}}{{{k_{22}}}} - {\eta _2}k_{11}^2\left( {{q^2} + k_{11}^2} \right)\frac{{\sin ({k_{12}}x)}}{{{k_{12}}}}} \right]\\
& - \frac{F}{{{\eta _1}{\eta _2}}}\left\{ {\left[ {1 - \left( {k_{11}^2 + k_{21}^2 - {q^2}} \right)\left( {{q^2} + 4{\kappa ^2}} \right)} \right]{q^2}\cos (qx) + 2q\kappa \left( {1 - {q^4} - 4{\kappa ^2}{q^2}} \right)\sin (qx)} \right\},
\end{align}
\end{subequations}
where $\eta_1=(q^2-k_{11}^2)^2+4q^2\kappa^2$, $\eta_2=(q^2-k_{21}^2)^2+4q^2\kappa^2$ with $\Phi_2$ and $\Psi_2$ are the solutions of homogenous system, given in (\ref{pfn}). The solutions in the regions 1 (metallic) and 2 (vacuum) can be matched via appropriate boundary conditions. Assuming, $c_1=c_2=c_4=0$ and $c_3=1$ without loss of generality, we arrive at the following continuous solutions in the region 1
\begin{subequations}\label{solpfm}
\begin{align}
&\Phi_1(x)=\frac{1}{\alpha }\left[ {\cos ({k_{11}}x) - \cos ({k_{21}}x)} \right],\\
&\Psi_1(x)= {1 + \frac{1}{\alpha }\left[ {k_{21}^2\cos ({k_{21}}x) - k_{11}^2\cos ({k_{11}}x)} \right]}.
\end{align}
\end{subequations}
with the normalized wavefunction defined as, $\Psi_{N1}(x)=\Psi_1(x)/\Psi_N$, where $\Psi_N$ is the normalizing factor, so that, $\int_{ - L}^0 {\left[ {\Psi _{N1}^2(x)} \right]dx = 1}$. The normalization factor, then, reads
\begin{subequations}\label{N}
\begin{align}
&{\Psi _N} = \left\{ {1 + \frac{{a_1^2 + a_2^2}}{2} + \frac{{a_1^2}}{{8L{k_{11}}}}\sin (2{k_{11}}L) + \frac{{a_2^2}}{{8L{k_{21}}}}\sin (2{k_{21}}L)} \right.\\
&{\left. { + \frac{{{a_1}}}{L}\left[ {\frac{1}{{{k_{11}}}} + \frac{{{a_2}{k_{11}}\cos ({k_{21}}L)}}{{k_{11}^2 - k_{21}^2}}} \right]\sin ({k_{11}}L) + \frac{{{a_2}}}{L}\left[ {\frac{1}{{{k_{21}}}} - \frac{{{a_1}{k_{21}}\cos ({k_{11}}L)}}{{k_{11}^2 - k_{21}^2}}} \right]\sin ({k_{21}}L)} \right\}^{1/2}},
\end{align}
\end{subequations}
where $a_1=(\alpha-E)/2\alpha$ and $a_2=(\alpha+E)/2\alpha$. The $d_i$ coefficients of wavefunctions in the spill-out region, are given as
\begin{subequations}\label{bc}
\begin{align}
&{d_1} = \frac{{F\left\{ {\alpha \left( {{q^4} + 1} \right) + k_{21}^2{\eta _1} - k_{11}^2{\eta _2} - {q^2}\left[ {{\eta _1} - {\eta _2} + \alpha \left( {k_{11}^2 + k_{21}^2 + 4{\kappa ^2}} \right)} \right]} \right\}}}{{2{\eta _1}{\eta _2}\left( {k_{11}^2 - k_{21}^2} \right)}},\\
&{d_2} = \frac{{F\kappa \left\{ {\alpha {k^4}\left( {3k_{11}^2 - 3k_{21}^2 - \alpha } \right) - \left( {k_{11}^2 - k_{21}^2 + \alpha } \right)\left[ {\left( {\alpha  + k_{21}^2{\eta _1}} \right) - k_{11}^2{\eta _2}} \right]} \right\}}}{{\alpha {\eta _1}{\eta _2}\left( {k_{11}^2 - k_{21}^2} \right)}}\\
& + \frac{{F\kappa \left( {k_{11}^2 - k_{21}^2 + \alpha } \right)\left\{ {2\alpha {\kappa ^4} - {k^2}\left[ {{\eta _1} - {\eta _2} + \alpha \left( {k_{11}^2 + k_{21}^2} \right)} \right]} \right\}}}{{\alpha {\eta _1}{\eta _2}\left( {k_{11}^2 - k_{21}^2} \right)}},\\
&d_3= - \frac{{\left( {k_{21}^2 - k_{11}^2 + \alpha } \right)}}{{2\left( {k_{11}^2 - k_{21}^2} \right)}}\\
& + \frac{{F\left\{ {k_{21}^2{\eta _1}\left( {{q^2} - k_{21}^2} \right) - k_{11}^2{\eta _2}\left( {{q^2} - k_{11}^2} \right) - {q^2}\alpha \left[ {1 - \left( {k_{11}^2 + k_{21}^2 - {q^2}} \right)\left( {{q^2} + 4{\kappa ^2}} \right)} \right]} \right\}}}{{2{\eta _1}{\eta _2}\left( {k_{11}^2 - k_{21}^2} \right)}},\\
&{d_4} = \frac{{2\kappa \left[ {\left( {k_{21}^2 + \alpha } \right) - k_{11}^2} \right]{\eta _1}{\eta _2} - F\kappa \left[ {k_{11}^4 + \alpha {q^4}\left( {k_{11}^2 + k_{21}^2 - 12{\kappa ^2}} \right)} \right]}}{{{\eta _1}{\eta _2}\left( {k_{11}^2 - k_{21}^2} \right)}},\\
& + \frac{{F\kappa \left\{ {k_{21}^4{\eta _1} + 3\alpha q - {q^2}\left[ {\alpha  + 3\left( {k_{21}^2{\eta _1} - k_{11}^2{\eta _2}} \right) + 4\alpha {\kappa ^2}\left( {k_{11}^2 + k_{21}^2} \right)} \right]} \right\}}}{{{\eta _1}{\eta _2}\left( {k_{11}^2 - k_{21}^2} \right)}}.
\end{align}
\end{subequations}
Ignoring the transient part of the solutions in region 2, we arrive at the following simple steady state solutions far beyond the metallic interface
\begin{subequations}\label{ntsol}
\begin{align}
&{\Phi_{ss}}(x) =  - A\sin (qx + \gamma),\hspace{3mm}\gamma  = \arctan \left[ {\frac{{\left( {{q^2} - k_{11}^2} \right)\left( {{q^2} - k_{12}^2} \right) - 4{q^2}{\kappa ^2}}}{{2\kappa q\left( {k_{11}^2 + k_{12}^2 - 2{q^2}} \right)}}} \right],\\
&{\Psi_{ss}}(x) = A\sin (qx + \delta),\hspace{3mm}\delta  = \arctan \left[ {\frac{{\left( {k_{11}^2 + k_{12}^2 - {q^2}} \right)\left( {{q^2} + 4{\kappa ^2}} \right) - 1}}{{2(\kappa /q)\left( {{q^4} + 4{\kappa ^2}{q^2} - 1} \right)}}} \right].
\end{align}
\end{subequations}
where $A=F/\eta_1\eta_2$ is the amplitude of the steady state solutions.

\begin{figure}[ptb]\label{Figure3}
\includegraphics[scale=0.6]{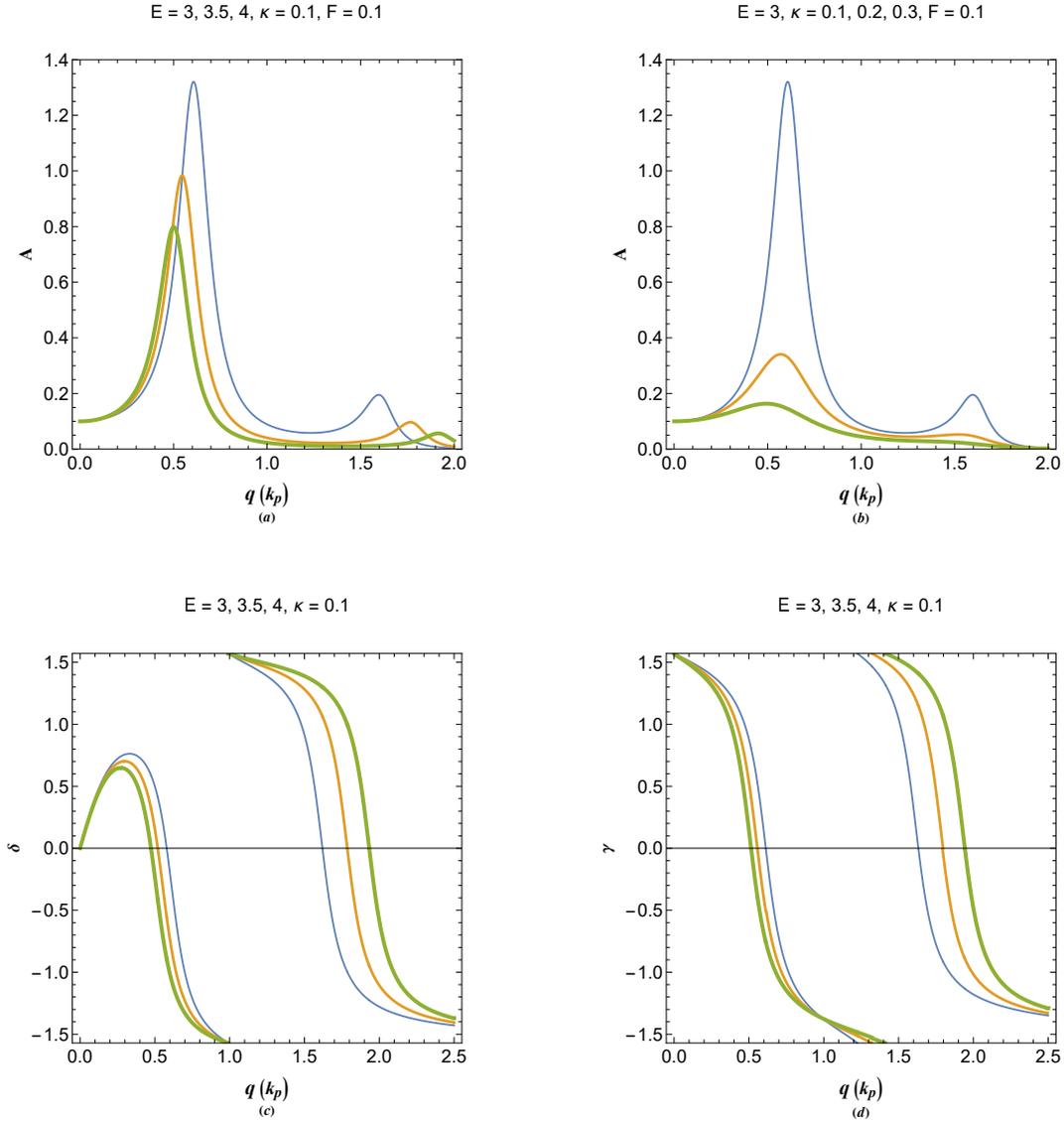}\caption{(a) The dual-resonance amplitude profiles in terms of the driving pseudoforce wavenumber for different values of the quasiparticle energy, $E$. (b) The resonance amplitude profiles in terms of the driving pseudoforce wavenumber for different values of the damping parametr, $\kappa$. (c) The phase angle $\delta$ between steady state wavefunction and the driving pseudoforce. (d) The phase angle $\gamma$ between the electrostatic potential energy and the driving pseudoforce. The increase in thickness of curves in plots refer to increases in the varied parameter above given frames.}
\end{figure}

Figure 3 depicts the amplitude of steady state solution as a function of the input wavenumber for different values of the normalized quasiparticle energy orbital, $E$. It is clearly remarked that there is a dual resonance feature occurring due to both wave-like and particle-like wavenumber match with that of the driving pseudoforce. However, the current plasmon resonance effect is caused by the wavenumber matching unlike conventional cases which takes place due to the frequency matching, hence, it is called the pseudo-resonance effect. Figure 3(a) shows that the increase in quasiparticle energy leads to further separation of the resonance peaks, whereas, the amplitude of both peaks decreases. Conversely, increase in the damping parameter leads to sharp decrease in peak magnitudes, whereas the their positions are left unaffected, as shown in Fig. 3(b). Figure 3(c) depicts the phase angle between the steady state solution and the driving pseudo-force as a function of the driving wavenumber. It is remarked that the angle changes sign at resonant wavenumbers, which is a known characteristic feature of the resonant coupled systems. The phase angle variation between the driving force and the electrostatic energy function is shown in Fig. 3(d). It is remarked that there is a phase lag between the wavefunction and electrostatic energy for small driving wavenumbers, whereas, for large wavenumbers no phase lag occurs.

\begin{figure}[ptb]\label{Figure4}
\includegraphics[scale=0.65]{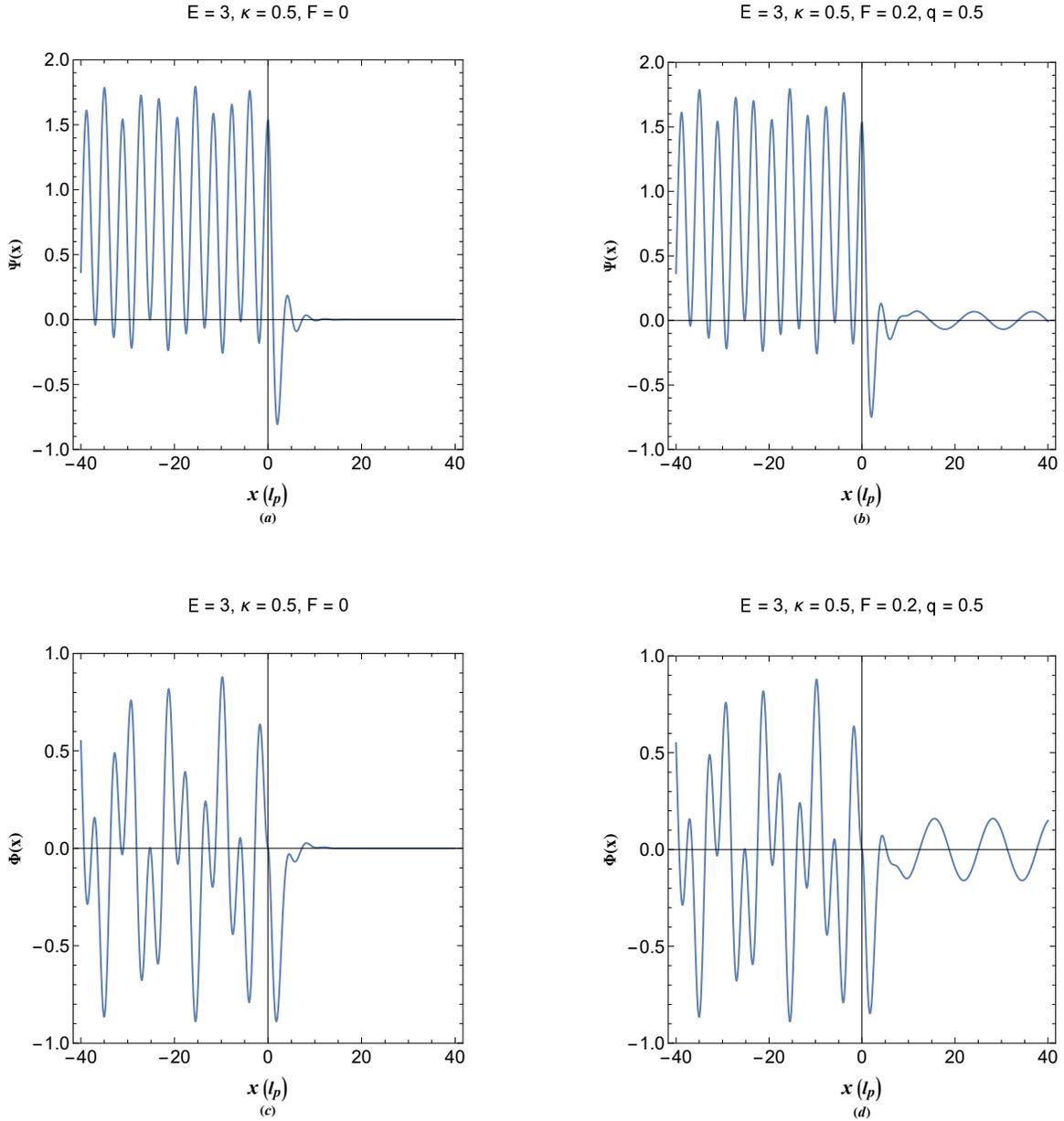}\caption{(a) The wavefunction profile in metallic and vacuum region for given parameters in the absence of driving pseudoforce. (b) The wavefunction profile in metallic and vacuum region for given parameters in the presence of driving pseudoforce. (c) The electrostatic energy profile in metallic and vacuum region for given parameters in the absence of driving pseudoforce. (d) The electrostatic energy profile in metallic and vacuum region for given parameters in the presence of driving pseudoforce.}
\end{figure}

Figure 4 shows the local wavefunction and electrostatic energy in the metallic and vacuum regions connected by the boundary conditions for given values of the energy orbital, $E$, damping parameter, $\kappa$ and pseudoforce magnitude, $F$. The dual-tone nature of oscillations is clearly apparent in the metallic region. In the absence of driving pseudoforce, the wavefunction in the spill-out region is of damped oscillatory type, as shown in Fig. 4(a). However, as remarked from Fig. 4(b), in the presence of pseudoforce, the oscillation extend further into the vacuum. Such plane-wave oscillations beyond the metallic interface may be an indication of resonant collective electron tunneling or amplified electromagnetic radiation with matched wavenumber of quasiparticle energy orbital and the corresponding driving psedoforce can be either an electron beam or the electromagnetic wave directed into the metallic surface at the resonant angle. Figure 4(c) shows the electrostatic energy distribution around the metal-vacuum interface. The oscillation in metallic region are again dual-tone and strongly damp in vacuum side. The Fig. 4(d) shows a similar feature as Fig. 4(b) for the case of electrostatic potential energy in the presence of driving pseudoforce.

The solutions to psedoforce system, studied sofar, represent the pure collective quantum states. The collective statistical quantities corresponding to an equilibrium state are, however, mixed quantum states which can be obtained via the standard averaging of the given quantity over all possible quasiparticle energy orbital using the appropriate occupation function and the Dos. For instance, the local electron distribution is given as
\begin{equation}\label{ns1}
\left\langle {{n}(x)} \right\rangle  = \frac{{\int\limits_2^\infty  {{{\left| {{\Psi_{N}}(x)} \right|}^2}g(E,\kappa )f(E,\theta )dE} }}{{\mathop \smallint \limits_2^\infty  g(E,\kappa )f(E,\theta )dE}},
\end{equation}
where the density of state in the plasmonic energy band is given in terms of the energy eigenvalues, $E$
\begin{equation}\label{ge}
g(E,\kappa ) = \frac{{\pi {{\left( {E + \alpha } \right)}^2}\sqrt {E - 2{\kappa ^2} + \alpha } }}{{\sqrt 2 \left[ {E\left( {E + \alpha } \right) - 4} \right]}}.
\end{equation}
On the other hand, the electrostatic energy distribution is
\begin{equation}\label{phi1}
\left\langle {\Phi(x)} \right\rangle  = \frac{{\int\limits_2^\infty  {\Phi(x)g(E,\kappa )f(E,\theta )dE} }}{{\int\limits_2^\infty  {g(E,\kappa )f(E,\theta )dE} }}.
\end{equation}

\begin{figure}[ptb]\label{Figure5}
\includegraphics[scale=0.6]{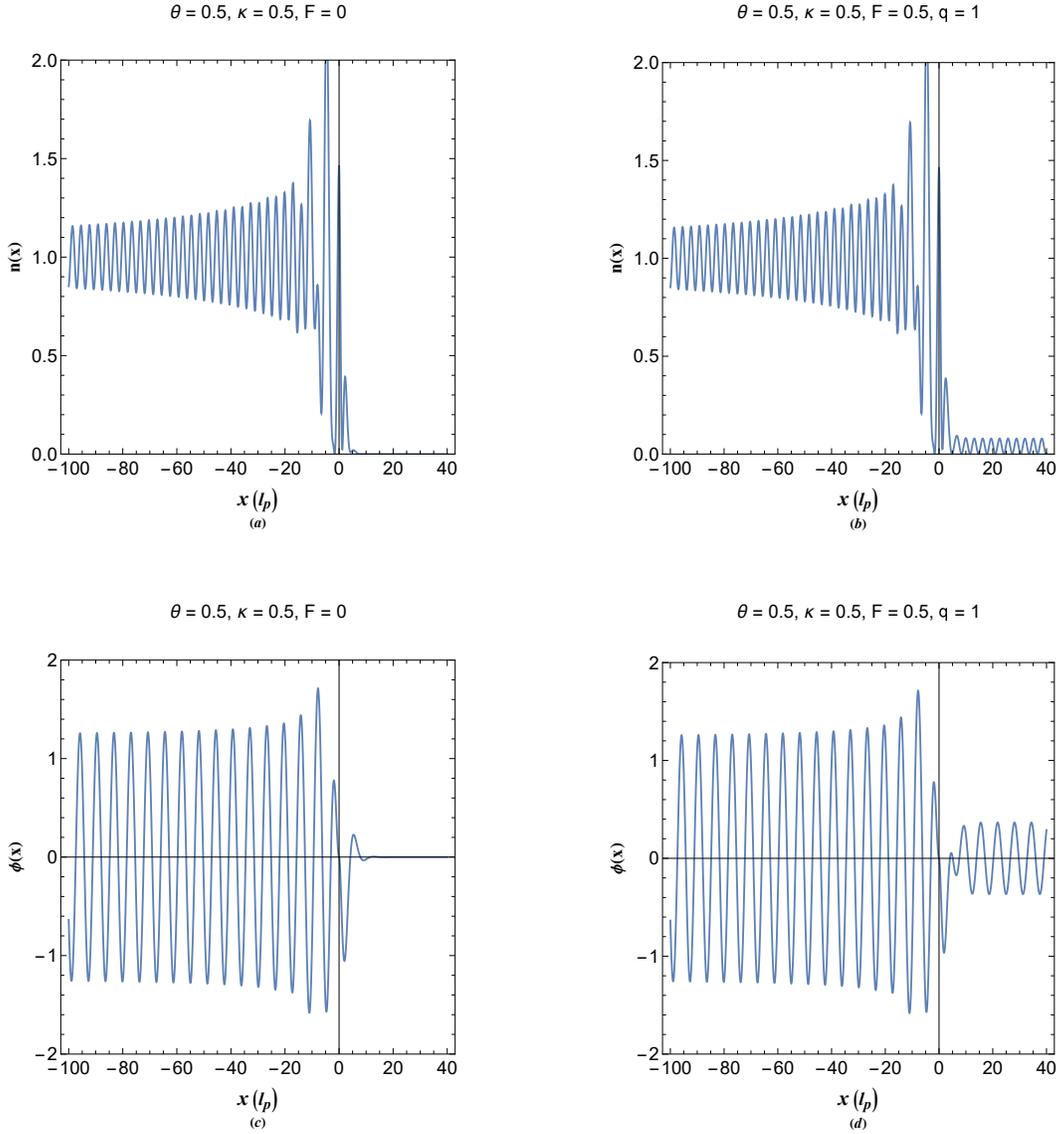}\caption{(a) The quantum statistically averaged electron number density distribution in metal-vacuum configuration for given normalized temperature and damping parameter, in the absence of external driving pseudoforce. (b) The quantum statistically averaged electron number density distribution in metal-vacuum configuration for given normalized temperature and damping parameter, in the presence of external driving pseudoforce. (c) The quantum statistically averaged electrostatic potential energy distribution in metal-vacuum configuration for given normalized temperature and damping parameter, in the absence of external driving pseudoforce. (d) The quantum statistically averaged electrostatic potential energy distribution in metal-vacuum configuration for given normalized temperature and damping parameter, in the presence of external driving pseudoforce.}
\end{figure}

Figure 5 shows the local equilibrium electron number density and electrostatic potential energy distribution in the metal-vacuum junction for given plasma parameters in the absence and presence of the driving pseudoforce. Figure 5(a) reveals that in the metallic region density distribution is strongly oscillatory with the amplitude of oscillations increasing toward the metallic edge. Before the metal edge, in the metallic region, a strong electron depletion occurs leading to a pronounced electric dipole, hence, an abrupt change in the dielectric properties is expected at the metal-vacuum interface. Beyond the metal interface, however, the sharp oscillatory decrease in the electron concentration ocuurs, in the absence of driving pseudoforce. Figure 5(b) shows the electron number density in the presence of the pseudoforce with given amplitude and driver wavenumber and similar other parameters as in Fig. 5(a). It is remarked that, presence of driving pseudoforce does not significantly affect the plasmon excitations in the metallic region. However, electron density of spill-out region extends to far beyond the interface. Note that amplitude of resonant spill-out electron excitations depends on the damping parameter and the temperature other than the driving force amplitude and wavenumber. Figure 5(c) depicts the spacial electrostatic energy distribution in the metallic and vacuum region in the absence of driving pseudoforce. Similar features as in Fig. 5(a) are present in Fig. 5(c) showing that the electrostatic potential energy drops sharply before the metallic edge due to the electron depletion. It is also remarked that the electrostatic energy penetrates into the vacuum even in the absence of a driving but quickly fades further away from the interface. The spill-out of electrons has the same oscillatory behavior as the static charge screening and can produce Friedel-like feature at the metallic edge. The later feature can have fundamental effect on scanning probe microscopy (SPM) measuring quantum device operations such as the scanning tunneling microscope (STM).

\begin{figure}[ptb]\label{Figure6}
\includegraphics[scale=0.6]{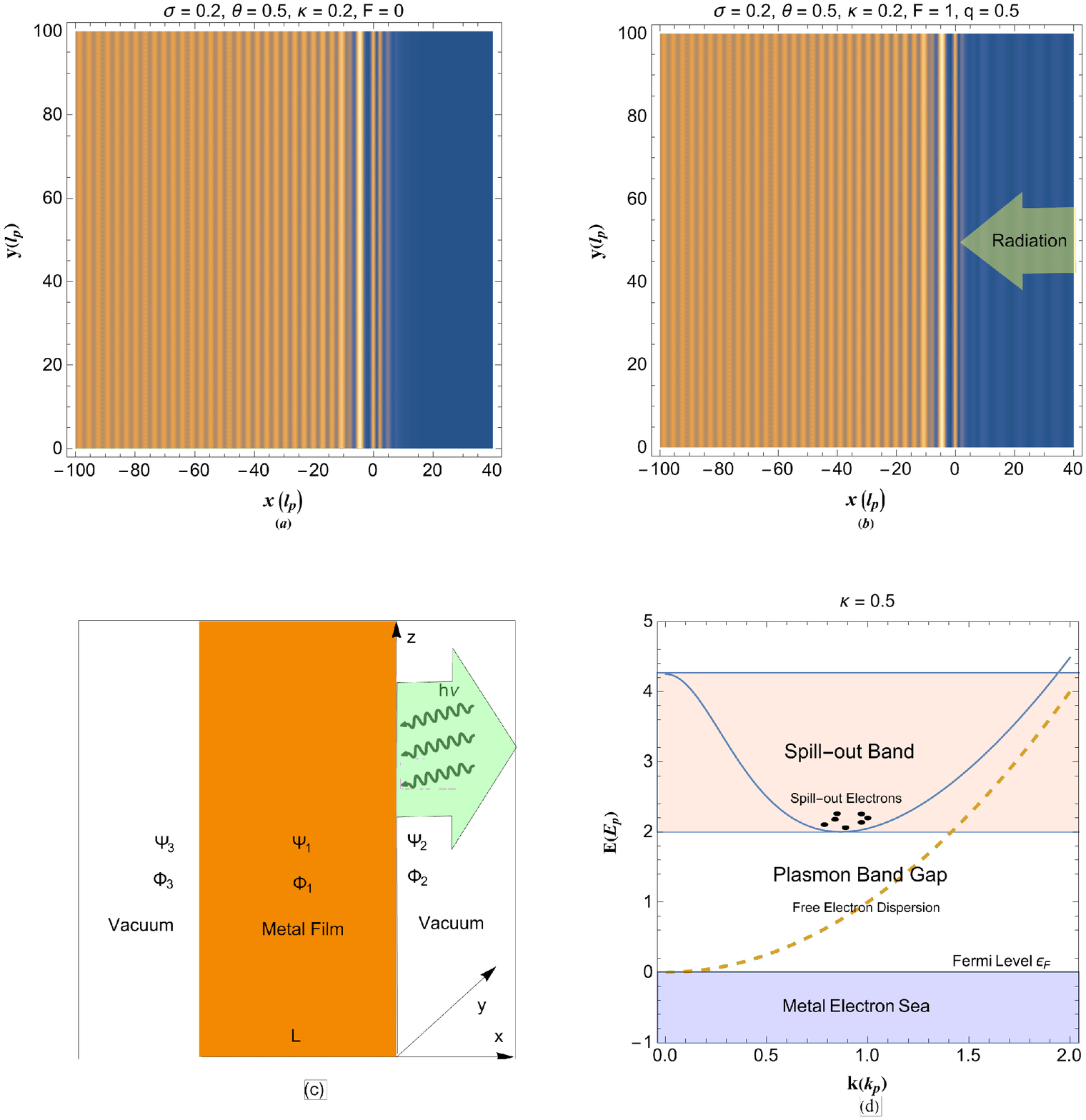}\caption{(a) The density plot of electron distribution around the metal-vacuum interface in the absence of driving pseudoforce showing the large pre-interface electric dipole. (b) The density plot of electron distribution around the metal-vacuum interface in the presence of driving pseudoforce for given values of normalized temperature, chemical potential, damping parameter and driving parameters. (c) The schematic plot of vacuum-metal-vacumm sandwich in metal slab configuration. (d) Dynamics of spill-out electrons in the damped plasmon energy band structure.}
\end{figure}

Figures 6(a) and 6(b) show the schematic density plot of the electron density distribution in the absence and presence of driving pseudoforce (which is assumed to be the radiation). In Fig. 6(a) the electron depletion has led to the formation of strong post interface dipole. The surface plasmon resonance is shown in Fig. 6(b) where the driving pseudoforce has been applied. It is remarked that the surface charge oscillations is significantly manipulated by the presence of pseudoforce. Figure 6(c) shows the schematic of the surface plasmon resonance in vacuum-metal-vacuum sandwich. Note that vacuum side may be replaced with a dielectric layer with the dielectric permittivity ratio $\epsilon_r$. In that case the plasmon frequency definition replaces with $\omega_p=\sqrt{4\pi e^2 n_0/(\epsilon_r m_e)}$. Each region in the plot is characterized by two functions corresponding to the region. This is contrary to the conventional quantum theory which uses only the wavefunction to characterize the single electron behavior in a given potential. The appropriately irradiated surface spill-out electrons cause the resonant electron-photon interactions and a collective tunneling or amplified reflected radiation intensity. Figure 6(d) shows the electron dynamics in the energy band structure of the plasmonic device. At an equilibrium temperature finite number of electrons are excited to the plasmon energy band in which collective oscillations among the excited electrons takes place. The plasmon band plays the role of a stirring room to produce hot electrons where the electrons behave collectively analogous to the starling murmuration effect in nature. Note that at low electron concentration or relatively higher temperatures greater majority of electron excite to the plasmon band. However, in the metallic dense electron regime only a fraction of electrons below the Fermi energy level take part in the collective phenomena. Therefore, the effective plasmon frequency of the quantum system, which is defined erroneously based on the Fermi electron density, must be reformulated according to the new equation of state (EoS) for plasmon electrons in an equilibrium temperature.

Statistical averages of other quantities, such as the (pseudo)resonance amplitude, $A_s=\left\langle {A} \right\rangle$, and the (pseudo)resonance power, $P_s$ can be obtained, accordingly. The pseudoresonance power may be written as
\begin{equation}\label{pow}
P = \frac{{{F^2}q}}{{{\eta _1}{\eta _2}}}\cos (qx)\cos (qx + \delta),
\end{equation}
Averaging over the complete cycle one gets
\begin{equation}\label{powa}
\bar P = \frac{{{F^2}q\cos\delta}}{{2{\eta _1}{\eta_2}}},
\end{equation}
where $\cos\delta$ is the pseudoresonance power coefficient given by
\begin{equation}\label{phase}
\cos\delta  = \frac{{q\left[ {\left( {k_{11}^2 + k_{21}^2 - {q^2}} \right)\left( {{q^2} + 4{\kappa ^2}} \right) - 1} \right]}}{{\sqrt {{q^2}{{\left[ {\left( {k_{11}^2 + k_{21}^2 - {q^2}} \right)\left( {{q^2} + 4{\kappa ^2}} \right) - 1} \right]}^2} + 4{\kappa ^2}{{\left( {{q^4} + 4{q^2}{\kappa ^2} - 1} \right)}^2}} }}.
\end{equation}
The quantum statistically averaged pseudoresonance power over all energy orbital leads to
\begin{equation}\label{pows}
\left\langle {\bar P} \right\rangle  = \frac{{\int\limits_2^\infty  {\bar P f(E,\theta )g(E,\kappa )dE} }}{{\int\limits_2^\infty  {f(E,\theta )g(E,\kappa )dE} }}.
\end{equation}

\begin{figure}[ptb]\label{Figure7}
\includegraphics[scale=0.6]{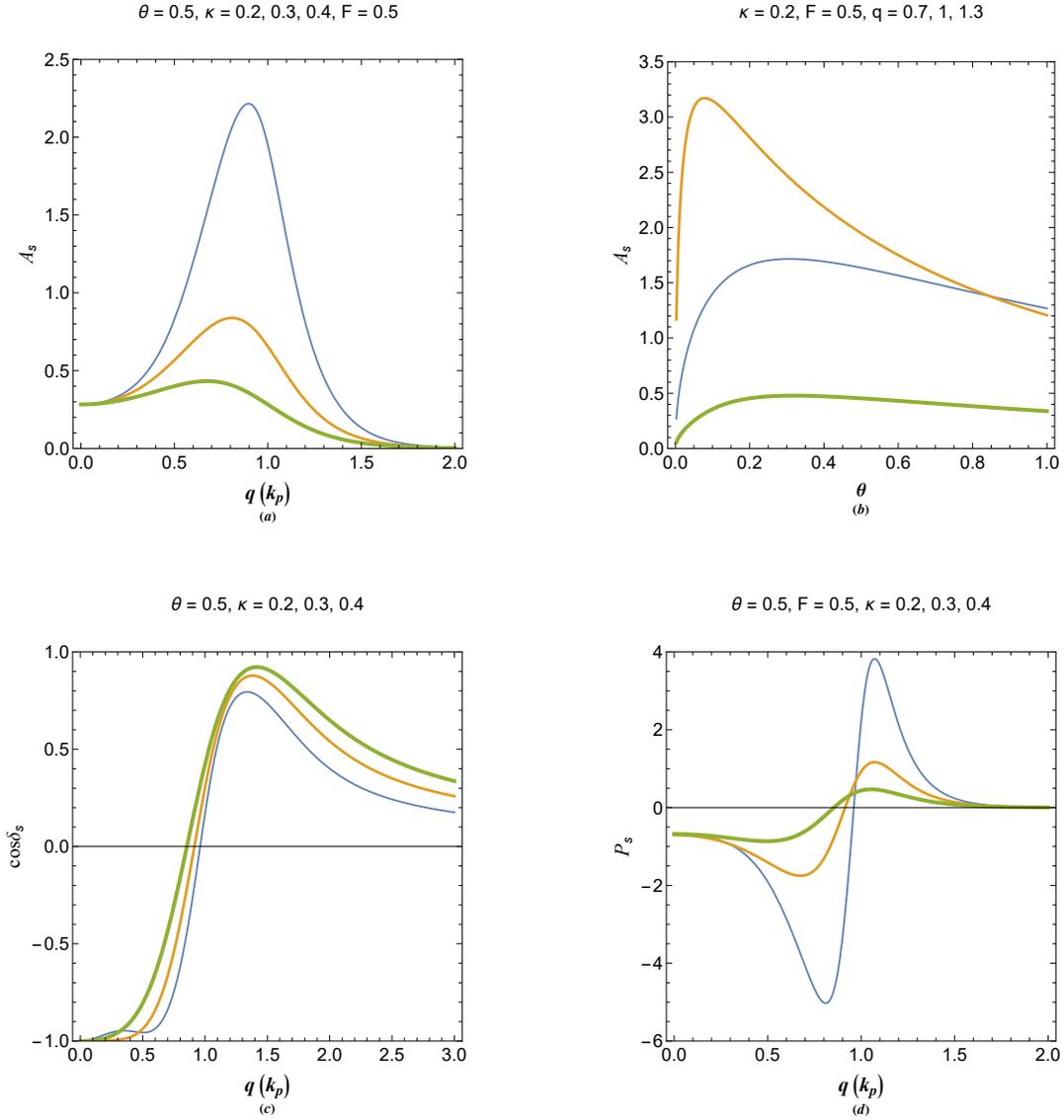}\caption{(a) The variation of psedoresonance amplitude with the driver wavenumber for different values of damping paramter. (b) Temperature dependence of the pseudoresonance amplitude for different values  of driver wavenumber. (c) The power coefficient of the pseudoresonance for various damping level. (d) The power loss/gain in pseudo resonance in terms of driver wavenumber for different damping values. The increase in thickness of curves in plots refer to increases in the varied parameter above given frames.}
\end{figure}

Figure 7 shows the variation in quantum statistically averaged pseudoresonance parameters of the driven spill-out electrons. The variation of steady state damped driven solution amplitude is depicted in Fig. 7(a) for different values of the damping parameter and fixed other parameters. It is remarked that the amplitude maximized close to the plasmon wavenumber for small damping. The amplitude decreases sharply and its maximum shifts toward smaller driving wavenumbers as the damping increases. The temperature dependence of the pseudoresonance amplitude is shown in Fig. 7(b). It is remarked that, close to the plasmon wavenumber driving the resonance amplitude maximizes at small temperatures. However, the later feature appears insignificant for wavenumbers further away from the critical wavenumber. The variation of power coefficient, $\cos\delta$ is shown with respect to the driving frequency for different damping parameter values in Fig. 7(c). This quantity changes the sign at the critical driving $q=1$, whereas, it maximizes at small $q$ and approximately $q=1.5$. It is also remarked that this quantity is not significantly affected by the change in damping parameter. In Fig. 7(d) the variation in pseudoresonance power is depicted for different damping parameter values. The quantity changes the sign close to the critical wavenumber indicating a power loss/gain below/above the corresponding critical point. It is remarked that the magnitude of power loss/gain strongly varies with the damping parameter.

\begin{figure}[ptb]\label{Figure8}
\includegraphics[scale=0.6]{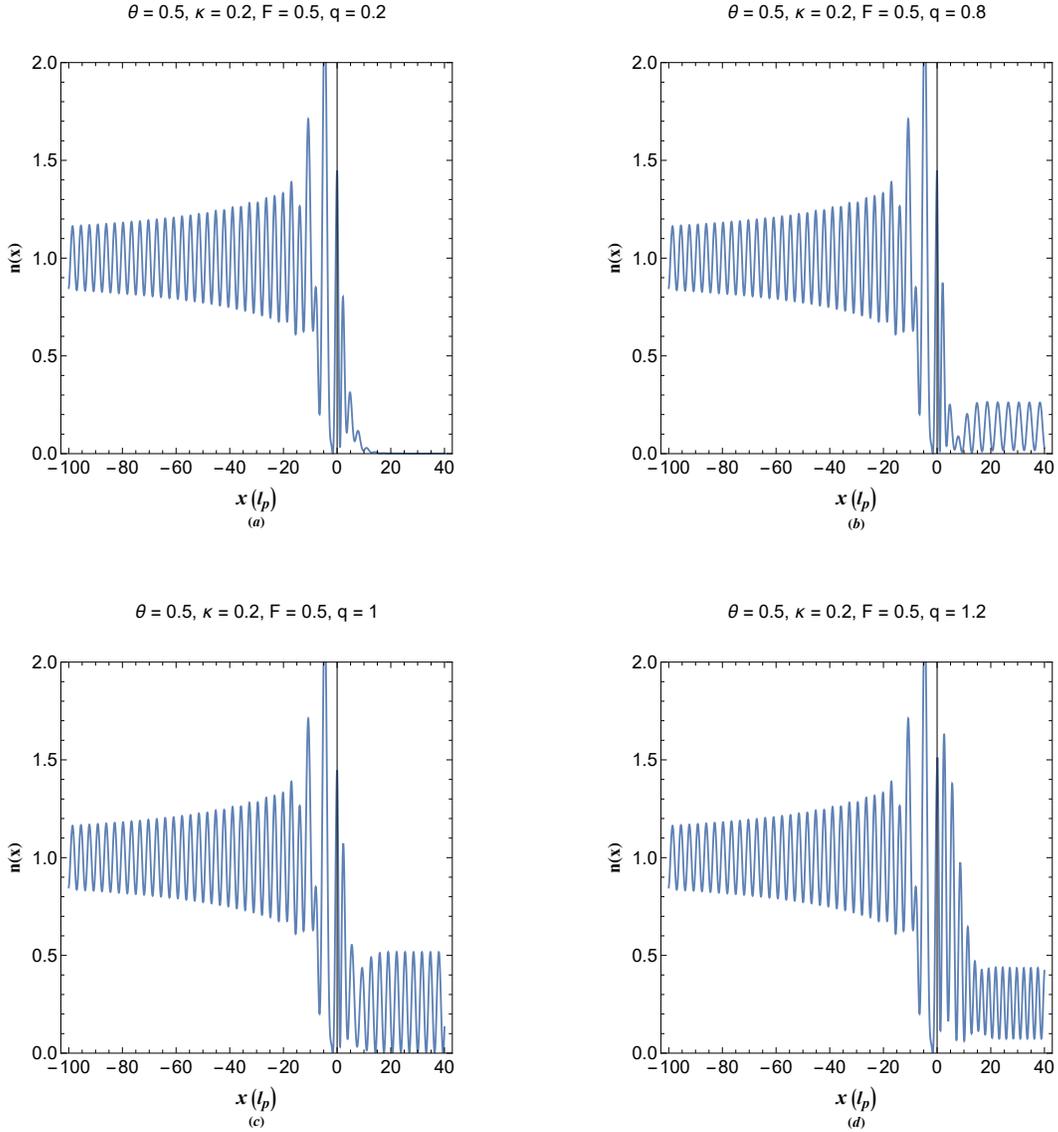}\caption{The electron number density distribution of driven surface plasmon resonance for different values the driving wavenumber and similar other parameters. (a) Small under-critical regime $q=0.2$. (b) Under-critical regime $q=0.8$. (c) Near-critical wavenumber regime $q=1$. (d) Over-critical wavenumber regime $q=1.2$.}
\end{figure}

The electron density profiles for different wavenumber regimes are compared in Fig. 8. It is remarked that the electron density profile close to the driven metallic interface strongly depends on the driver wavenumber regime. For small wavenumbers $q=0.2$, in Fig. 8(a), the resonance amplitude is very small with the transient state extending to average distances beyond the interface. For $q=0.8$ the same feature repeats with increased resonance amplitude. Close to the critical wavenumber the transient surface density has increased significantly. Finally, for overcritical wavenumber value $q=1.2$ the electron accumulation in front of the metallic surface is further increased. It seems that the sign of the pseudoresonance power has a keen relation to the momentum transfer to the surface electrons.

\begin{figure}[ptb]\label{Figure9}
\includegraphics[scale=0.6]{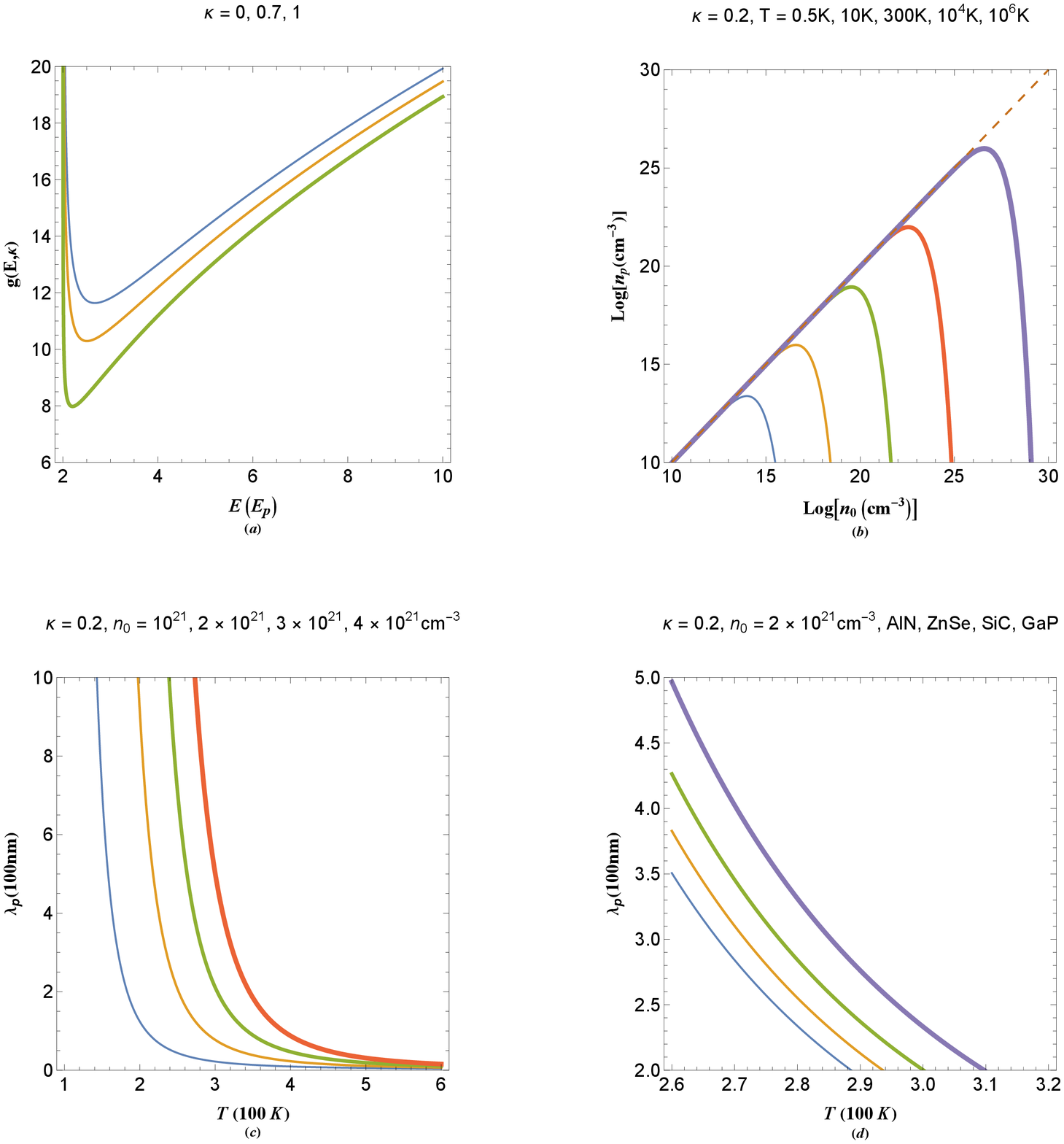}\caption{(a) The plasmon energy density of states for different values of the damping parameter. (b) The plasmon electron number density versus the total electron number density of the electron gas for different values of the electron temperatures. (c) Variation of the plasmon wavelength $\lambda_p=2\pi/k_p$ with respect to temperature for different values of the total electron number density. (d) Variation of the plasmon wavelength for different dielectric replaces with the vacuum. The dashed line in (b) indicates the plasmon saturation limit $n_p=n_0$. The increase in thickness of curves in plots refer to increases in the varied parameter above given frames.}
\end{figure}

In Fig. 9 we show some important quantities regarding the DoS of plasmon energy band, the EoS of plasmon electrons, the real plasmon wavelength defined in terms of number density of the excited electron to the plasmon band. As already mentioned, the plasmon scale parameter $l_p$ depicted in Fig. 1 is defined through the total electron concentrations in the metal. However, the correct definition must be via the EoS of the excited electrons. The energy DoS of the plasmon band as given in Eq. (\ref{ge}) is plotted in Fig. 9(a). It is seen that the DoS diverges for minimum plasmon conduction energy, $E=2$, and has minimum value which shifts toward this critical value as the damping parameter increases. However, the increase in damping parameter leads to overall decrease of plasmon DoS. Note that for large energies the DoS approaches the free electron values $g_0(E)=2\pi\sqrt{E_0}$ with $E_0=k^2$ is the normalized free electron energy dispersion. The number density of plasmon electrons is shown in Fig. 9(b) in terms of the total electron density, for different values electron temperature. The dashed line in (b) indicates the plasmon saturation limit $n_p=n_0$. The plasmon electron density is obtained using the following conventional definition
\begin{equation}\label{eosp}
{n_p}(\sigma,\theta,\kappa) = {n_0(\sigma,\theta)}\frac{{\int_2^\infty  {g(E,\kappa)f(E,\theta )dE} }}{{\int_0^\infty  {{g_0}(E_0){f}(E_0,\theta )dE_0} }},
\end{equation}
where $n_0={n_c}{\theta ^6}{\rm{L}}{{\rm{i}}_{3/2}}[-{\rm{exp(}}\sigma {\rm{/}}\theta {\rm{)}}]^4$ is the total electron density given by (\ref{np}) with $n_c=16{e^6}{m^3}/{\pi ^3}{\hbar ^6}$ being the characteristic electron number density, $\rm{Li}$ is the polylog function and $f(E,\theta)=1/[1+\exp(E/\theta)]$, is the plasmon electron occupation function. It is remarked that for a given electron temperature up to a lower critical value of, $n_0$, we have $n_p=n_0$, that is, all the electrons in the gas contribute to the collective phenomena, whereas, with increase of $n_0$ beyond this critical value the $n_p$ sharply falls off and vanishes for another upper critical, $n_0$. Hence for a given temperature there is a cutoff plasmon electron density and a maximum value before this cutoff value. The later phenomenon is an important feature of the quantum plasmon excitations which is usually ignored in the plasmonic research. The cutoff electron density increases with increase of the electron gas temperature. Variations of plasmon wavelength $\lambda_p=2\pi/k_p$ (in nanometer unit) is depicted in Fig. 9(c) with respect to total electron temperature (in $100$K unit) for different total electron number density values. It is shown that for larger electron concentrations the plasmon wavelength increases significantly at lower temperatures and sharply decreases at higher temperatures. The variation of $\lambda_p$ with the temperature is shown for the case where the given semiconductor is replaced by the vacuum side. The dielectric values are $\epsilon_r(AlN)\simeq 4.53$, $\epsilon_r(ZnSe)\simeq 5.4$, $\epsilon_r(SiC)\simeq 6.7$ and $\epsilon_r(GaP)\simeq 9.09$, respectively in the plot from thin to thick curves.

\begin{figure}[ptb]\label{Figure10}
\includegraphics[scale=0.6]{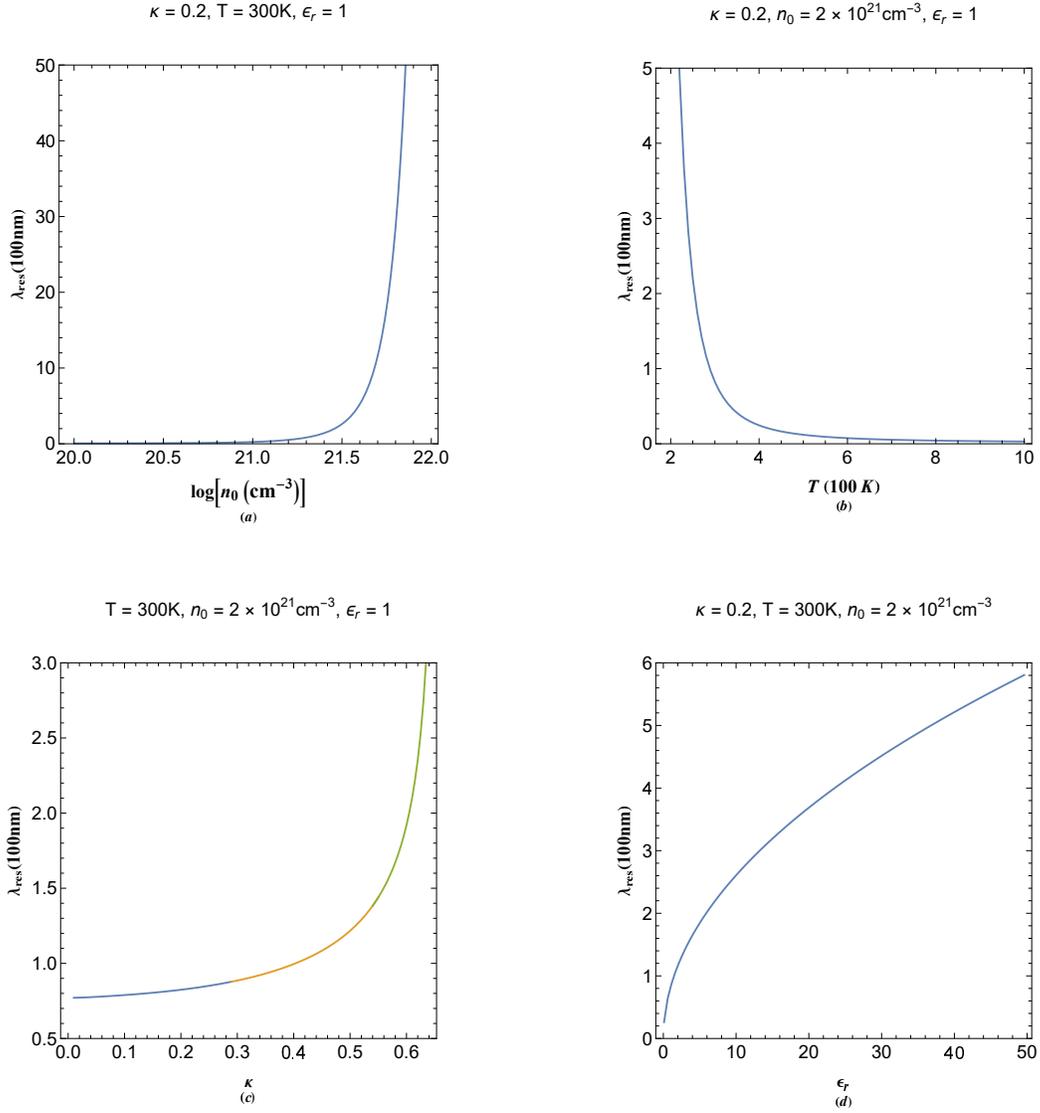}\caption{(a) The dependence of the resonace wavelength on total free electron density for given temperature and damping parameter. (b) Dependence of the resonance wavelength on the electron temperature for given electron temperature and dampind parameter. (c) Dependence of the resonance on the damping parameter for given electron density and temperature. (d) Dependence of the resonance wanelength in the relative dielectric permittivity of subsatnce replaced with the vacuum.}
\end{figure}

In Fig. 10 we have shown the variations of the surface plasmon resonance wavelength in nanometers in terms of various electron gas parameters. Figure 10(a) shows that the resonance wavelength increases sharply with the total electron number density of electrons falling in the visible range for electron density $10^{19}$cm$^{-3}$$<n_0<10^{20}$cm$^{-3}$ at room temperature and $\kappa=0.2$. Figure 10(b) reveals that $\lambda_{res}$ decreases sharply with increase of the electron temperature for given electron concentration. In Fig. 10(c) variation in resonance wavelength is studied in terms of the damping parameter for given values of electron density and temperature. It is remarked that increase in this parameter leads to increase in the resonance wavelength. Finally, Fig. 10(d) reveals that increase of the dielectric permittivity of substance replaced with the vacuum leads to increase in the resonance wavelength for given values of other parameters. Note that in current model of the surface plasmon resonance we have assumed that the spill-out electrons have the same normalized temperature as the metal electrons. This is far from being exact because the plasmon temperature depends on the plasmon electron density which itself is calculated from the normalized temperature. Therefore, for correct theoretical modeling the plasmon spill-out electron density can be measured, directly, using Langmuir probe and then a better estimate for the plasmon energy of spill-out region can be obtained.

\begin{figure}[ptb]\label{Figure11}
\includegraphics[scale=0.6]{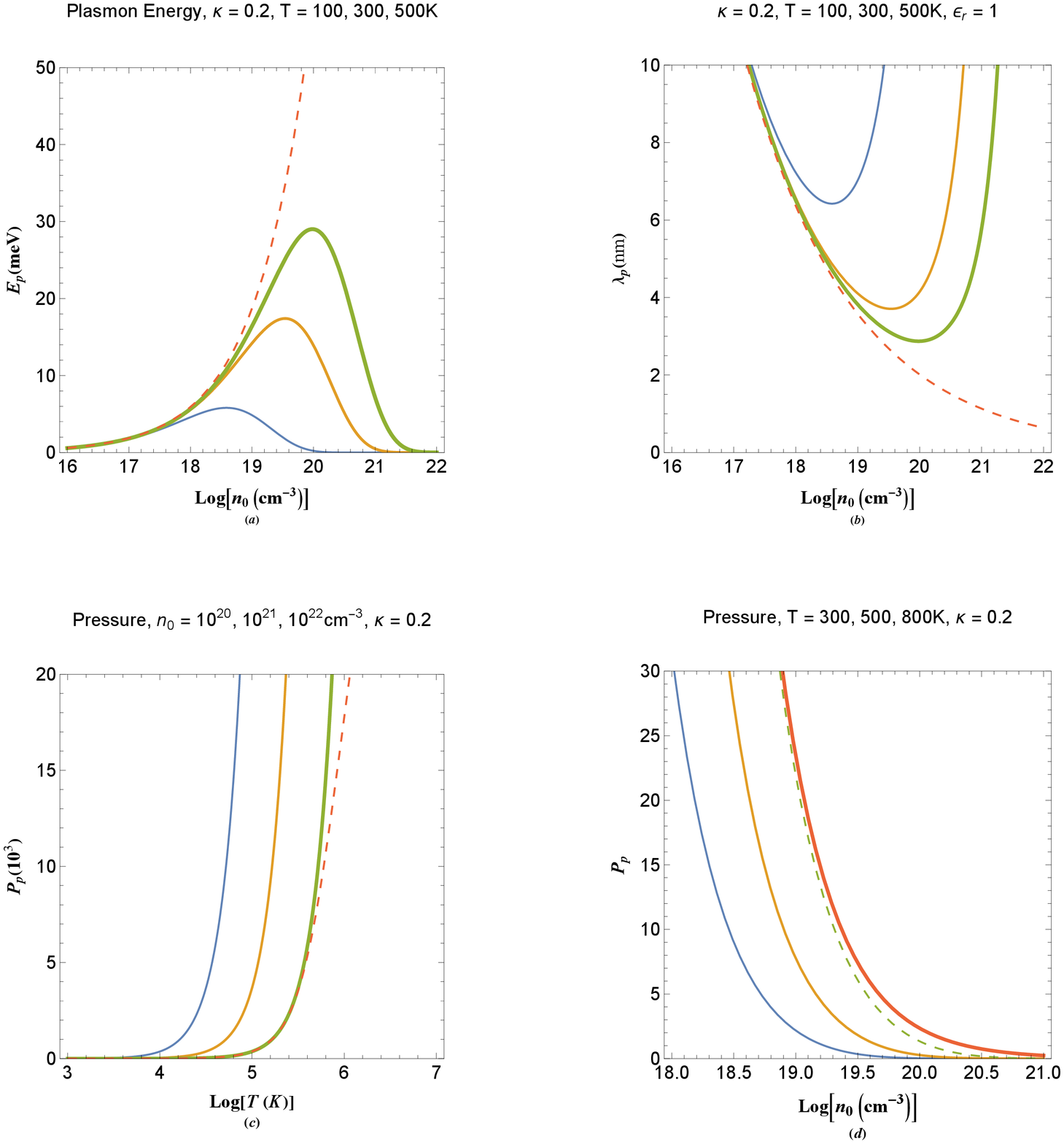}\caption{(a) The variation of plasmon energy with respect to the total number density of electron gas as calculated based on the excited electron number density. (b) The variation of plasmon wavelength (resonance scale-length) with respect to the total number density of electron gas as calculated based on the excited electron number density. (c) The plasmon pressure in terms of electron temperature for different total electron number density. (d) The plasmon pressure in terms of total electron number density for different electron temperature. The dashed curves correspond to free electron energy dispersion. The increase in thickness of curves in plots refer to increases in the varied parameter above given frames.}
\end{figure}

Finally, in Fig. 11 we have shown the characteristic (not conventional) energy and length scales variation with the total number density of electron gas calculated based on the excited electrons to the plasmon energy band for given electron temperature. The dashed curves correspond to the free electron dispersion relation. It is remarked from Fig. 11(a) that the plasmon energy has a maximum value depending of the electron temperature and the maximum plasmon energy shifts to higher electron density as the temperature increases. The overall plasmon energy value also increases with increase of the electron temperature. It is noted that the plasmon energy of free electron gas (dashed curve) increases monotonically. Figure 11(b) reveals that the plasmon wavelength, which is a characteristic scale of surface plasmon resonance effect, has a minimum value shifting higher electron densities as the temperature increases. This minimum also decreases substantially with increase of electron temperature. The plasmon length of free electron gas (dashed curve) decreases monotonically. Figure 11(c) shows the plasmon electron pressure in terms of the electron temperature for different total electron concentrations. It can be shown that for large values of the temperature the plasmon pressure obeys a simple relation of the form, $P_p= T +  10^\tau$, $\tau\simeq 1.1122$. It is also remarked that increase in the total electron density lowers the plasmon pressure and that the free electron pressure corresponding to the largest value (dashed curve) is even lower. Figure 11(d) depict the plasmon pressure as a function of total electron number density for different electron temperatures. It is shown that the plasmon pressure of an arbitrary degenerate electron gas is higher for lower total electron number density and higher temperatures. Pressure of the free electron gas (dashed curve) indicates a slightly lower pressure corresponding to $T=800$K.

\section{Conclusion}

In this research we developed a theory of surface resonance effect based on the effective Schr\"{o}dinger-Poisson system. The Hermitian and non-Hermitian models were used to respectively, model the metallic and spill-out electron excitations which are shown to be of dual-scale type. The quantum statistically averaged resonance parameters were obtained and discussed in terms of various plasmon parameters. Particularly, it was shown that the conventional definitions of plasmon parameter should be corrected for dense electron gas based on the electron transition probability into the plasmon energy band. It was shown that the resonance wavelength falls into the visible and infrared region for typical lower degenerate electron number densities at room temperature. Current theory can further elucidate the underlying physical mechanism involved in collective quantum resonance and unveils the dominant effect of electrostatic interactions in collective phenomena.

\section{Data Availability}

The data that support the findings of this study are available from the corresponding author upon reasonable request.


\section{References}

\begin{thebibliography}{}
\bibitem {atw} H. A. Atwater, The Promise of Plasmonics, Sci. Am. \textbf{296}, 56(2007); doi.org/10.1038/scientificamerican0407-56
\bibitem{man1} G. Manfredi, Preface to Special Topic: Plasmonics and solid state plasmas, Phys. Plasmas \textbf{25}, 031701(2018); https://doi.org/10.1063/1.5026653
\bibitem{maier} S. A. Maier, Plasmonics: Fundamentals and Aplications, Springer Science Business Media LLC (2007).
\bibitem{xu} Xu, X., Li, H., Hasan, D., Ruoff, R. S., Wang, A. X. and Fan, D. L., Near-Field Enhanced Plasmonic-Magnetic Bifunctional Nanotubes for Single Cell Bioanalysis, Adv. Funct. Mater. \textbf{23} 4332 (2013);  doi:10.1002/adfm.20120382
\bibitem{haug} H. Haug and S. W. Koch, "Quantum theory of the optical and electronic properties of semiconductors", World Scientific, 2004,
\bibitem{mark} P. A. Markovich, C.A. Ringhofer, and C. Schmeister, Semiconductor Equations (Springer, Berlin, 1990).
\bibitem {seeger} K. Seeger, Semiconductor Physics (Springer, Berlin, 2004) 9th ed.
\bibitem {hu} C. Hu, Modern Semiconductor Devices for Integrated Circuits (Prentice Hall, Upper Saddle River, New Jersey, 2010) 1st ed.
\bibitem{yofee} A. D. Yofee, Low-dimensional systems: quantum size effects and electronic properties of semiconductor microcrystallites (zero-dimensional systems) and some quasi-two-dimensional systems, Adv. Phys., \textbf{42}, 173-262(1993), DOI: 10.1080/00018739300101484
\bibitem{zhu} Zhu, Xiaoli; Gao, Tao, Li, Genxi (ed.), "Nano-Inspired Biosensors for Protein Assay with Clinical Applications", Elsevier, (2019), ISBN 978-0-12-815053-5
\bibitem{jian} Jianing Chen, Michela Badioli, Pablo Alonso-González, Sukosin Thongrattanasiri, Florian Huth, Johann Osmond, Marko Spasenović, Alba Centeno, Amaia Pesquera, Philippe Godignon, Amaia Zurutuza Elorza, Nicolas Camara, F. Javier García de Abajo, Rainer Hillenbrand, Frank H. L. Koppens, Optical nano-imaging of gate-tunable graphene plasmons, Nat. \textbf{487} 77(2012); doi.org/10.1038/nature11254
\bibitem{fey1} Z. Fei, A. S. Rodin, G. O. Andreev, W. Bao, A. S. McLeod, M. Wagner, L. M. Zhang, Z. Zhao, M. Thiemens, G. Dominguez, M. M. Fogler, A. H. Castro Neto, C. N. Lau, F. Keilmann, D. N. Basov, Gate-tuning of graphene plasmons revealed by infrared nano-imaging, Nat. \textbf{487} 82(2012); doi:10.1038/nature11253
\bibitem{hugen} Hugen Yan, Tony Low, Wenjuan Zhu, Yanqing Wu, Marcus Freitag, Xuesong Li, Francisco Guinea, Phaedon Avouris, Fengnian Xia, Damping pathways of mid-infrared plasmons in graphene nanostructures, Nat. Photonics \textbf{7} 394(2013); doi:10.1038/nphoton.2013.57
\bibitem{sun1} Xijiao Mu and Mengtao Sun, Interfacial charge transfer exciton enhanced by plasmon in 2D in-plane lateral and van der Waals heterostructures, Appl. Phys. Lett. \textbf{117}, 091601 (2020); doi.org/10.1063/5.0018854
\bibitem{sun2} Rui Yang, Yuqing Cheng and Mengtao Sun, Aluminum plasmon-enhanced deep ultraviolet fluorescence resonance energy transfer in h-BN/graphene heterostructure, Optics Comm. \textbf{498}, 127224 (2021); doi.org/10.1016/j.optcom.2021.127224
\bibitem{sun3} Jianuo Fan, Jizhe Song, Yuqing Cheng and Mengtao Sun, Pressure-dependent interfacial charge transfer excitons in WSe2-MoSe2 heterostructures in near infrared region, Results in Physics \textbf{24}, 104110 (2021); doi.org/10.1016/j.rinp.2021.104110
\bibitem {and} P. Andrew and W. L. Barnes, Energy transfer across a metal film mediated by surface plasmon polaritons. Science \textbf{306}, 1002 (2004).
\bibitem {stock} M. I. Stockman, Nanofocusing of optical energy in tapered plasmonic waveguides. Phys. Rev. Lett. \textbf{93}, 137404 (2004).
\bibitem {qui} M. Quinten, A. Leitner, J. R. Krenn and F. R. Aussenegg, Electromagnetic energy transport via linear chains of silver nanoparticles. Opt. Lett. \textbf{23}, 1331 (1998).
\bibitem {muh} P. Mühlschlegel, H. J. Eisler, O. J. F. Martin, B. Hecht and D. W. Pohl, Resonant optical antennas. Science \textbf{308}, 1607 (2005).
\bibitem {umm} S. Ummethala, T. Harter, K. Koehnle, et al., THz-to-optical conversion in wireless communications using an ultra-broadband plasmonic modulator, Nat. Photonics \textbf{13}, 519(2019); doi.org/10.1038/s41566-019-0475-6
\bibitem {zhuo} Z. Zhou, E. Sakr, Y. Sun, P. Bermel, Solar thermophotovoltaics: reshaping the solar spectrum. Nanophotonics \textbf{5} 1 (2016).
\bibitem {goy} I. Goykhman, B. Desiatov, J. Khurgin, J. Shappir, U. Levy, Locally oxidized silicon surface-plasmon Schottky detector for telecom regime. Nano Lett \textbf{11} 2219 (2011).
\bibitem {haf} C. Haffner, D. Chelladurai, Y. Fedoryshyn, et al., Low-loss plasmon-assisted electro-optic modulator. Nature \textbf{556} 483 (2018).
\bibitem {kar} A. V. Krasavin and N. I. Zheludev, Active plasmonics: Controlling signals in Au/Ga waveguide using nanoscale structural transformations. Appl. Phys. Lett.
\textbf{84}, 1416 (2004).
\bibitem {oul} R. F. Oulton, et al. Plasmon lasers at deep subwavelength scale. Nature \textbf{461}, 629 (2009).
\bibitem {cesar} C. Calvero, Plasmon-induced hot-electron generation at nanoparticle/metal-oxide interfaces for photovoltaic and photocatalytic devices, Nat. Photonic \textbf{8}, 95(2014); doi.org/10.1038/nphoton.2013.238
\bibitem {jac} Jacob B. Khurgin, Fundamental limits of hot carrier injection from metal in nanoplasmonics, Nanophotonics \textbf{9(2)}, 453(2020); doi.org/10.1515/nanoph-2019-0396
\bibitem {at2} H. A. Atwater and A. Polman, Plasmonics for improved photovoltaic devices, Nat. Mater. \textbf{9}, 205(2010); DOI: 10.1038/nmat2629
\bibitem {kit} C. Kittel, Introduction to Solid State Physics, (John Wiely and Sons, New York, 1996), 7th ed.
\bibitem {ash} N. W. Ashcroft and N. D. Mermin, Solid State Physics (Saunders College Publishing, Orlando, 1976).
\bibitem{fetter} Alexander L. Fetter, Edge magnetoplasmons in a bounded two-dimensional electron fluid, Phys, Rev. B, \textbf{32} 7676 (1985).
\bibitem{mahan} G. D. Mahan, Many-particle physics, 2nd edition, chapter 5 (Plenum press, New York, 1990).
\bibitem{fey} Feynman, Richard, QED: The Strange Theory of Light and Matter, Princeton University Press (1985).
\bibitem {fermi} E. Fermi and E. Teller, The Capture of Negative Mesotrons in Matter, Phys. Rev. {\bf 72}, 399 (1947).
\bibitem {bohm} D. Bohm and D. Pines, Coulomb Interactions in a Degenerate Electron Gas, Phys. Rev. \textbf{92} 609(1953).
\bibitem {bohm1} D. Bohm, Avoiding Negative Probabilities in Quantum Mechanics, Phys. Rev. \textbf{85}, 166–193 (1952).
\bibitem {bohm2} D. Bohm, A Suggested Interpretation of the Quantum Theory in Terms of "Hidden" Variables, Phys. Rev. \textbf{85}, 180–193 (1952).
\bibitem {pines} D. Pines, A Collective Description of Electron Interactions, Phys. Rev. \textbf{92} 626 (1953).
\bibitem {levine} P. Levine and O. V. Roos, Plasma Theory of the Many-Electron Atom, Phys. Rev, \textbf{125} 207(1962).
\bibitem {klim} Y. Klimontovich and V. P. Silin, Plasma Physics, edited by J. E. Drummond (McGraw-Hill, New York, 1961).
\bibitem{5} T. Takabayasi, On the Formulation of Quantum Mechanics associated with Classical pictures, Prog. Theor. Phys., \textbf{8} 143(1952).
\bibitem{6} T. Takabayasi, Relativistic Hydrodynamics of the Dirac Matter. Part I. General Theory, Prog. Theor. Phys., \textbf{14} 283 (1955).
\bibitem{7} T. Takabayasi, On the Hydrodynamical Representation of Non-Relativistic Spinor Equation, Prog. Theor. Phys., \textbf{12} 810 (1954).
\bibitem{8} T. Takabayasi, On the Separability of Dirac Equation, Prog. Theor. Phys., \textbf{9} 681 (1953).
\bibitem{9} T. Takabayasi, Relativistic particle with internal rotational structure, Nuovo Cim, \textbf{13} 532 (1959).
\bibitem{50} C. Castro and E. Zuazua, Flux identification for 1-d scalar conservation laws in the presence of shocks, Math. Comp. \textbf{1} 38 (2006).
\bibitem{51} C. Castro, Nonlinear corrections to the Schr\"odinger equation from geometric quantum mechanics, J. Math. Phys., \textbf{31} 2633 (1990).
\bibitem{52} D. L. Johna, L. C. Castro, and D. L. Pulfrey, Quantum capacitance in nanoscale device modeling, J. Appl. Phys., \textbf{96} 5180 (2004).
\bibitem{53} C. Castro, On Weyl geometry, random processes, and geometric, Found. Phys., \textbf{80}(276) 2025 (2011).
\bibitem{lind} N. Bohr, J. and Lindhard, Electron Capture and Loss by Heavy Ions Penetrating through Matter, Dan. Mat. Fys. Medd. \textbf{28}, 1 (1954).
\bibitem{bonitz1} M. Bonitz, E. Pehlke, and T. Schoof, Phys. Rev. E \textbf{87}, 033105 (2013).
\bibitem{sea1} P. K. Shukla, B. Eliasson, and M. Akbari-Moghanjoughi, Phys. Rev. E \textbf{87}, 037101 (2013).
\bibitem{bonitz2} M. Bonitz, E. Pehlke, and T. Schoof, Phys. Rev. E \textbf{87}, 037102 (2013).
\bibitem{sea2} P. K. Shukla, B. Eliasson and M. Akbari-Moghanjoughi, Phys. Scr. \textbf{87} 018202 (2013).
\bibitem{bonitz3} M. Bonitz, E. Pehlke, and T. Schoof, Phys. Scr. \textbf{88}, 057001 (2013).
\bibitem{akbarihd} M. Akbari-Moghanjoughi, Phys. Plasmas \textbf{22}, 022103 (2015); {\it ibid.} {\bf 22}, 039904 (E) (2015).
\bibitem{sm} L. G. Stanton and M. S. Murillo, Phys. Rev. E \textbf{91}, 033104 (2015); {\em ibid.} \textbf{91}, 049901 (E) (2015).
\bibitem{michta} D. Michta, F. Graziani, and M. Bonitz, Contrib. Plasma Phys. \textbf{55}, 437 (2015).
\bibitem{moldabekov} Zh. Moldabekov, T. Schoof, P. Ludwig, M. Bonitz, and T. Ramazanov, Phys. Plasmas \textbf{22}, 102104 (2015).
\bibitem{gard} C. Gardner, The quantum hydrodynamic model for semiconductor devices, SIAM, J. Appl. Math. \textbf{54} 409(1994).
\bibitem{ichimaru1} S. Ichimaru, Strongly coupled plasmas: high-density classical plasmas and degenerate electron liquids, Rev. Mod. Phys. {\bf 54}, 1017 (1982); doi.org/10.1103/RevModPhys.54.1017
\bibitem{ichimaru2} S. Ichimaru, H. Iyetomi, and S. Tanaka, Statistical physics of dense plasmas: Thermodynamics, transport coefficients and dynamic correlations, Phys. Rep. {\bf 149}, 91 (1987); doi.org/10.1016/0370-1573(87)90125-6
\bibitem{ichimaru3} S. Ichimaru, Statistical Physics: Condensed Plasmas, Addison Wesely, New York, 1994.
\bibitem {man0} G. Manfredi, How to model quantum plasmas, Fields Inst. Commun. {\bf 46}, 263–287 (2005); Proceedings of the Workshop on Kinetic Theory (The
Fields Institute, Toronto, Canada 2004): http://arxiv.org/abs/quant--ph/0505004.
\bibitem {haas1} F. Haas, Quantum Plasmas: An Hydrodynamic Approach, Springer, New York, 2011.
\bibitem {moldbon} Zh. A. Moldabekov, M. Bonitz, and T. S. Ramazanov, Theoretical foundations of quantum hydrodynamics for plasmas, Phys. Plasmas \textbf{25}, 031903 (2018); https://doi.org/10.1063/1.5003910
\bibitem{kohn} W. Kohn, L. J. Sham, "Self-Consistent Equations Including Exchange and Correlation Effects". Physical Review. \textbf{140} (4A) 1133 (1965); doi:10.1103/PhysRev.140.A1133
\bibitem {elak} B. Eliasson and M. Akbari-Moghanjoughi, Phys. Lett. A, \textbf{380}, 2518(2016); doi.org/10.1016/j.physleta.2016.05.043
\bibitem{akblind} Akbari-Moghanjoughi, M. Plasmonic Dielectric Response of Finite Temperature Electron Gas, arXiv:1912.08769 (unpublished results)
\bibitem {markl} M. Marklund and G. Brodin, Dynamics of Spin-1/2 Quantum Plasmas, Phys. Rev. Lett. \textbf{98}, 025001(2007); doi.org/10.1103/PhysRevLett.98.025001
\bibitem {shuk} P. K. Shukla, B. Eliasson, Nonlinear aspects of quantum plasma physics, Phys. Usp. \textbf{53} 76(2010)
\bibitem {brd} L. Stenflo, G. Brodin, Large amplitude circularly polarized waves in quantum plasmas, J. Phys. Plasmas, \textbf{76} 261(2010).
\bibitem {se} P. K. Shukla and B. Eliasson, Nonlinear Interactions between Electromagnetic Waves and Electron Plasma Oscillations in Quantum Plasmas, Phys. Rev. Lett. \textbf{99}, 096401(2007).
\bibitem {sten} L. Stenflo, Resonant three-wave interactions in plasmas, Phys. Scr. \textbf{T50} 15(1994).
\bibitem {ses} P. K. Shukla, B. Eliasson, and L. Stenflo, Stimulated scattering of electromagnetic waves carrying orbital angular momentum in quantum plasmas, Phys. Rev. E \textbf{86}, 016403(2012).
\bibitem {brod1} G. Brodin and M. Marklund, Spin magnetohydrodynamics, New J. Phys. \textbf{9}, 277(2007).
\bibitem {mark1} M. Marklund and G. Brodin, Dynamics of Spin-$1/2$ Quantum Plasmas, Phys. Rev. Lett. \textbf{98}, 025001(2007).
\bibitem {man3} N. Crouseilles, P. A. Hervieux, and G. Manfredi, Quantum hydrodynamic model for the nonlinear electron dynamics in thin metal films, Phys. Rev. B {\bf 78}, 155412 (2008).
\bibitem{kim} Hwa-Min Kim and Young-Dae Jung, Landau damping effects on collision-induced quantum interference in electron-hole plasmas, EPL \textbf{79} 25001(2007).
\bibitem {fhaas} F. Haas, G. Manfredi, P. K. Shukla, and P.-A. Hervieux, Breather mode in the many-electron dynamics of semiconductor quantum wells, Phys. Rev. B, \textbf{80}, 073301 (2009).
\bibitem{akbquant} M. Akbari-Moghanjoughi, Quantized plasmon excitations of electron gas in potential well, Phys. Plasmas, \textbf{26}, 012104 (2019); doi.org/10.1063/1.5078740
\bibitem{akbheat} M. Akbari-Moghanjoughi, Heat capacity and electrical conductivity of plasmon excitations, Phys. Plasmas, \textbf{26}, 072106 (2019); doi.org/10.1063/1.5097144
\bibitem{akbdual} M. Akbari-Moghanjoughi, Effect of quantum charge screening on dual plasmon scattering, Phys. Plasmas, \textbf{26}, 112102 (2019); doi.org/10.1063/1.5123621
\bibitem{akbedge} M. Akbari-Moghanjoughi, Quantum edge plasmon excitations and electron spill-out effect, Phys. Plasmas \textbf{29}, 082112 (2022); doi.org/10.1063/5.0102151
\bibitem{akbint} M. Akbari-Moghanjoughi, Quantum interference of three dimensional plasmon excitations, Phys. Plasmas, \textbf{26}, 062105 (2019); doi.org/10.1063/1.5090366
\bibitem{akbnat1} M. Akbari-Moghanjoughi, Energy band structure of multistream quantum electron system, Sci. Rep. \textbf{11}, 21099 (2021).
\bibitem{akbnat2} M. Akbari-Moghanjoughi, Photo-plasmonic effect as the hot electron generation mechanism. Sci. Rep. \textbf{13}, 589 (2023); doi.org/10.1038/s41598-023-27775-1
\bibitem {haasbook} F. Haas, Quantum Plasmas: An Hydrodynamic Approach, Springer, New York, 2011.
\end{thebibliography}
\end{document}